\documentclass[reprint, 
preprintnumbers,amsmath,amssymb]{revtex4}
\usepackage{graphicx}
\usepackage{xcolor}

\begin{document}
\title{Non-Clausius heat transfer: The method of the nonstationary Langevin equation}
\author{Alex V. Plyukhin} 
\email{aplyukhin@anselm.edu} \affiliation{
Saint Anselm College, Manchester, New Hampshire, USA} 
\date{\today}

\begin{abstract}
Compared to other formulations of the second law of thermodynamics,
the Clausius statement that heat does not  spontaneously flow from cold to hot concerns a   
system in non-equilibrium states, and in that respect is more ambitious but also more ambiguous.
We discuss two scenarios when the Clausius statement in its plain form does not hold.
First, for ergodic systems, the energy transfer may be consistent with the statement
on a coarse-grained time scale, but  be anomalously directed during time intervals shorter 
than the thermalization time.
In particular, when an  initially colder system is brought in contact to a hotter bath,
the internal energy of the former increases with time in a long run but not monotonically. 
Second, the heat transfer may not respect the Clausius statement on any time-scale
in non-ergodic systems due to the formation of localized vibrational modes.
We illustrate the two scenarios with a familiar model of an isotope atom attached
to a semi-infinite harmonic atomic chain.
Technically, the discussion is based on a Langevin equation for the isotope,
using the initial condition when the isotope and chain
are initially prepared in uncorrelated canonical states under  the constraint
that the boundary atom between the isotope and chain is initially
fixed and later released.
In such setting, the noise in the Langevin equation is non-stationary,
and the fluctuation-dissipation relation has a non-standard form.

\end{abstract}


\maketitle
\section{Introduction}
Of many formulations of the second law of thermodynamics (the book~\cite{CS} counts 21 of them),  
the most versatile
one  is the Clausius inequality 
$\Delta S\ge \frac{\Delta Q}{T}$.
It establishes the low bound for the change of entropy $\Delta S=S_2-S_1$ of an open system which passes
from equilibrium state $1$ to equilibrium state $2$ as a result of 
receiving the amount of heat $\Delta Q$ from the environment at temperature $T$.
Transition $1\to 2$ may be either reversible (in which case $\Delta S=\Delta Q/T$) or non-reversible
(then $\Delta S>\Delta Q/T$), but since $S$ is defined only for equilibrium states,
the second law in the form of the Clausius inequality assumes that the initial and final states are equilibrium ones.

On the other hand, there are other formulations of the second law where
the equilibrium nature of initial and final state is not mentioned explicitly,
and as a matter of fact  is not assumed.
In particular, the Clausius statement reads as follows~\cite{C}:
``No process is possible whose sole result is the transfer of heat from a
  body of lower temperature to a body of higher temperature". 
In short, heat does not spontaneously flow from cold to hot.
This statement does not elaborate the nature of initial and final states. 
Applied literally to a system in a thermal contact with a hotter (colder) bath,
the Clausius statement implies that the internal energy of the system $U(t)$ increases
(decreases) monotonically until the system reaches thermal equilibrium with the bath.
The monotonicity is 
essential here because the Clausius statement tacitly implies that the derivative $U'(t)$ 
is of the same sign at any time, namely,
$U'(t)>0$ if the system is initially colder, and $U'(t)<0$ if the system is initially
hotter than the bath.

Clearly, the two formulations cannot be equivalent. The Clausius inequality is applied to two equilibrium states,
while the Clausius statement about the heat flow's direction implies no such restrictions. 
The application ranges of the two formulations
are overlapping but not identical. 
For instance, 
when  two semi-infinite systems of different temperatures are brought into
a thermal contact, our expectation
about the direction of the heat flow is based on the Clausius statement,
not on the Clausius inequality. The latter cannot be applied 
here (except perhaps when the temperature difference is infinitesimally small)
because  the overall combined system of infinite size does not reach thermal equilibrium on a finite time scale.

It therefore may appear that the Clausius statement is more general formulation of the second
law than the Clausius inequality. However, the Clausius statement has restrictions of its own.
In this paper, we consider a specific model of a microscopic system in contact with an infinite
bath and  show that the  Clausius statement may be violated in the following two scenarios.

{\it Scenario 1} assumes that the system is ergodic in the sense that eventually
it comes to thermal equilibrium with the bath. Suppose the system is initially colder than the bath.
Then we show that the system's internal  energy  $U(t)$ increases with time but not monotonically.
There are time intervals, albeit microscopically  short, when the derivative of $U(t)$
is negative, i.e., the colder system temporary releases heat into the hotter bath.
However, on a larger time scales, i.e., for sufficiently long time intervals $t_2-t_1>0$,
the internal energy's change  is positive,  $U(t_2)-U(t_1)>0$, in accordance  with the Clausius statement.

{\it Scenario 2} assumes that the system is nonergodic in the sense  
that it does not reach thermal equilibrium with the bath. 
For example, a light isotope atom does not reach equilibrium
with a uniform harmonic atomic chain due to formation of the localized vibrational mode (see below).
In that case, the system, which is initially colder than the bath, reaches a stationary
but not equilibrium state, in which its average 
over time energy may be lower than in the initial state. In other words,
the colder system may permanently release heat into a hotter bath.  

Both scenarios manifestly contradict the Clausius statement, involving heat transfer
from a colder system to a hotter one. We refer to this phenomenon as the non-Clausius heat transfer. 
In {\it Scenario 1}  the non-Clausius  heat transfer is transient, in {\it Scenario  2} it is permanent.
On the other hand, both scenarios  involve  a system
in nonequilibrium final states 
and  therefore do not violate the second law in the form of the Clausius inequality.

For a similar reason our discussion has no direct connection to
the  fluctuation theorem for 
heat exchange by Jarzynski and W\'ojcik~\cite{JW}. The theorem establishes the ratio of probabilities of system's
trajectories corresponding to Clausius and non-Clausius heat transfers
(i.e., for trajectories with the same amount but opposite signs of absorbed heat), but as the Clausius inequality,
the theorem assumes 
that the initial and final states are equilibrium ones. 
Also, the fluctuation theorem by Jarzynski and W\'ojcik is proved under the assumption of weak coupling between
the system and thermal bath. In contrast, in this paper we consider
a small system strongly coupled to the environment.

The possibility of anomalously directed  heat transfer was recently discussed in the literature
from different 
perspectives, both general and system specific~\cite{Gross,Hilbert1,Hilbert2,Swendsen,Hou,Lutz,Plyukhin},
sometimes with conflicting conclusions. In recent work~\cite{Plyukhin}, we discussed non-Clausius heat transfer
within familiar and exactly solvable Rubin's model~\cite{Weiss,Rubin} where the system is an isotope
atom embedded in the otherwise homogeneous harmonic chain.
Rubin's model and its modifications have been exploited in very many studies but, with only
a few exceptions~\cite{Hynes1,Hynes2,Bez}, with a very special initial condition. That condition,
which  is commonly used in microscopic derivations of Langevin and Fokker-Planck equations~\cite{MO,Weiss,Hanggi},
implies that 
at $t<0$ the isotope (system) is fixed and the chain (bath)
is equilibrated in the field of the fixed system.
At $t=0$, the system is released and instantaneously acquires
a desirable initial distribution, for instance the equilibrium canonical distribution with temperature $T_0$.
The latter can be interpreted as the initial temperature of the system. 
Considering that at $t<0$ the bath is correlated to the system,
the instantaneous change of the state of the system alone at $t=0$ appears to be a rather artificial assumption.

In this paper, we consider a model similar to that of Ref.~\cite{Plyukhin} except that
at $t<0$ we fix not the system but the system-bath boundary. 
This has the advantage that at $t<0$ 
both the system and bath are mobile and have
an opportunity (by means of coupling to  external reservoirs) 
to thermalize and acquire 
uncorrelated initial canonical distributions with given (in general different)
temperatures. Such  setting appears to be more natural, at least conceptually, in the context of
the heat transfer problem compared to standard Langevin models 
where the bath's initial distribution is correlated to the position of the initially fixed system, and the system
acquires a desirable initial distribution
instantaneously at $t=0$. 

The aforementioned modification, while it may appear only incremental, significantly reshapes
the theory and alters some predictions. Within the presented model
it is still possible (as in the model of Ref.~\cite{Plyukhin}) to describe the system
by the generalized Langevin equation,  but now it involves a non-stationary noise
related to the dissipation kernel via a non-standard fluctuation-dissipation relation.
The Langevin equation with a non-stationary noise, which we refer to for brevity
as the {\it non-stationary Langevin equation}, emerges
naturally in many fields, particularly for 
the description of open systems
interacting with nonequilibrium thermal bath ~\cite{SO,Evans,H,VH,K,Cui,Me,Visco,Maes}.
New phenomena in non-equilibrium environments came into limelight in recent years,
noticeably diffusion in living cells~\cite{Visco,Maes}.  In this paper we have a situation
when the bath is initially in equilibrium, but not in equilibrium with the system.
For that peculiar yet quite generic case we derive the generalized Langevin equation
with  a non-stationary noise but with a stationary (depending only on the time difference)  dissipation kernel.
The fluctuation-dissipation relation we obtain for that
case seems to be not covered by other models discussed in the literature.
Although in this paper the themes of non-Clausius heat transfer and of a non-stationary Langevin equation
are intertwined, the latter is  of interest of its own.

\section{Model}
For weakly coupled macroscopic systems
the expression ``to place system $A$ in a thermal contact to system $B$" does not involve any ambiguity. 
In contrast, for strongly coupled (small) systems such placing in general requires non-negligible  
mechanical work, which affects the systems' initial energy distribution. As a result, the initial condition
is determined not only by initial temperatures of the two systems, but also by the specific protocol 
according to which the systems  are brought into a physical contact.

\begin{figure}[t]
\includegraphics[height=5.5cm]{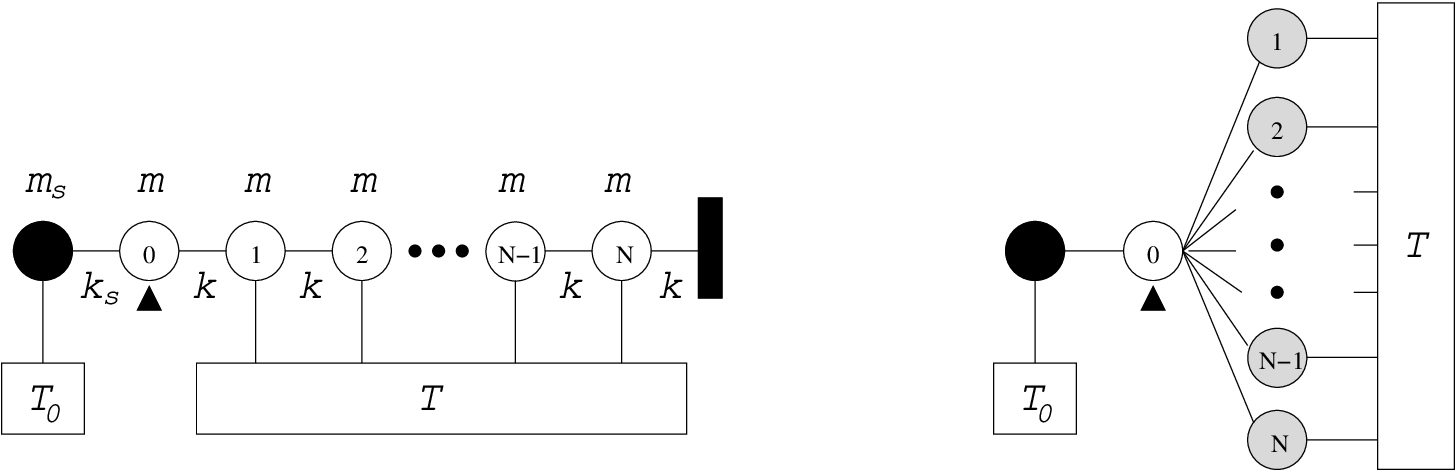}
\caption{ Left: The system under consideration at times $t<0$. The  isotope (black circle) 
is connected to an external thermal reservoir (smaller rectangle) with temperature $T_0$. 
The chain atoms (white circles) are connected to an external thermal reservoir (larger rectangle) 
of temperature $T$. The symbols $\blacktriangle$ indicates that boundary atom $i=0$ 
is fixed in its mechanical equilibrium positions blocking
the heat exchange between the isotope and the chain. 
As a result of such setting, at $t<0$ the isotope and the chain are prepared 
in uncorrelated canonical states with temperatures $T_0$ and $T$, respectively.
At the moment $t=0$ the coupling to external reservoirs
and the constraint $\blacktriangle$ are removed,
and for $t>0$ the overall system 
evolves as an isolated one.
Right: The same system but the part of the chain consisting of 
atoms $i=1\cdots N$ is represented as 
$N$ independent oscillators, or normal modes (gray circles).
}
\end{figure}

In the microscopic theory of Brownian motion is is usually assumed that the system of interest and thermal bath
are initially prepared according to the following protocol: At $t<0$ the system is 
fixed and the bath evolves in a potential created by the fixed system and reaches the constrained equilibrium. 
In this protocol 
the bath's initial distribution  at $t=0$ is developed as a result of the natural dynamical evolution, 
while the system, immediately after it is released  at $t=0$,
instantaneously acquires an arbitrary
initial distribution assigned ``by hand".  Because of the latter feature, we may refer to this 
protocol as ``sudden preparation".  
The advantage of the sudden preparation protocol is its simplicity, in particular, in the derivation 
and applications of the Langevin equation and fluctuation-dissipation relation.  
The disadvantage is an artificial way of assigning the initial condition  for the system.  
It is often not clear what physical setup, if any, can be responsible for a given initial distribution of the system.

In this paper we wish to overcome
the above-mentioned disadvantage 
of the sudden preparation protocol 
modifying   it 
in the following way:
At $t<0$, instead of  fixing  the system, we fix the position of the boundary between the system and bath. 
In such setting, not only the bath, but also the system is allowed to evolve naturally at $t<0$. 
With an additional assumption that at $t<0$  the system and bath are connected to external thermal 
reservoirs with given temperatures, this protocol  allows a more natural way to assign initial 
canonical distributions for the system and bath. 
As was mentioned in Introduction, 
this improvement comes with a price that  a Langevin equation for the system involves 
a non-stationary noise and the fluctuation-dissipation relation has a more complicated form.

We consider an isotope atom (or adatom) of mass $m_s$ attached by the linear spring with the stiffness
constant $k_s$ to the left end of the harmonic chain consisting of $N+1$ atoms of mass $m$ connected
by springs with the stiffness constant $k$, see Fig. 1. 
The isotope plays the role of a system of interest (hence the subscript $s$ in $m_s$ and $k_s$),
while the chain is an idealization of the thermal bath. We shall use the terms ``system"
and ``isotope", as well as ``bath" and ``chain", interchangeably.
The two parameters of the model are the mass ratio $\alpha$ and the ratio
of stiffness constants $\beta$,
\begin{eqnarray}
\alpha=\frac{m}{m_s}, 
\qquad \beta=\frac{k_s}{k}.
\label{alphabeta}
\end{eqnarray}
Comparing to a more familiar model  of an isotope in a uniform chain
and characterized by a single mass ratio parameter $\alpha$ (such model is often referred to as
Rubin's model~\cite{Weiss}), the presence of the second parameter $\beta$ offers  more flexibility.
In   particular, the model with two parameters (\ref{alphabeta})  gives a broader condition of
the localized mode formation, the phenomenon we shall find important in the present context.
The atoms of the chain are labeled by index $i=0 \cdots N$. The limit $N\to\infty$ will be eventually taken.
The right boundary atom of the chain $i=N$ is attached to the wall for all time by the same spring as for
the rest of the chain. The left boundary atom of the chain $i=0$ is fixed in its
mechanical equilibrium position for $t<0$ and released at $t=0$.

The model also involves implicitly two external thermal baths whose role is to prepare the  system (isotope)
and bath (chain) in states described by uncorrelated canonical distributions with given
temperatures $T_0$ and $T$, respectively. 
Therefore we assume that at $t<0$ the system is connected to the external thermal bath with
temperature $T_0$, and the bath  to another external bath with temperature $T$.
At $t=0$ the connection to external baths and the constraint on the boundary atom $i=0$
are removed, and the overall system (the isotope plus chain) evolves as an isolated one.

For $t>0$, i.e. for the stage of the unconstrained free evolution,  the Hamiltonian of the overall system is 
\begin{eqnarray}
H=\frac{p^2}{2m_s}+\sum_{i=0}^N \frac{p_i^2}{2m}+\frac{k_s}{2}\,(q-q_0)^2+
\frac{k}{2}\,(q_0-q_1)^2+\cdots+ \frac{k}{2}\,(q_{N-1}-q_N)^2+\frac{k}{2}\,q_N^2,
\label{H}
\end{eqnarray}
where $(q,\,p)$ and $\{q_i,\, p_i\}$ are coordinates and momenta of the isotope and chain's atoms,
respectively. As usual, as coordinates we choose  displacements of atoms from their mechanical equilibrium positions.

For $t<0$, i.e. for the stage of the constrained initial preparation, the boundary atom $i=0$
is fixed, so that $q_0=p_0=0$, and the Hamiltonian (\ref{H}) takes the form
\begin{eqnarray}
H\to H'=H_s+H_b,
\end{eqnarray}
where $H_s$ is the Hamiltonian of the system (isotope) in the field of 
the fixed boundary atom $i=0$,
\begin{eqnarray}
H_s=\frac{p^2}{2m_s}+\frac{k_s}{2}\,q^2,
\label{H_s}
\end{eqnarray}
and $H_b$ is the Hamiltonian of the bath (chain) with fixed boundary atom $i=0$, 
\begin{eqnarray}
H_b=\sum_{i=1}^N \frac{p_i^2}{2m}+
\frac{k}{2}\,q_1^2+\frac{k}{2}\,(q_1-q_2)^2\cdots+ \frac{k}{2}\,(q_{N-1}-q_N)^2+\frac{k}{2}\,q_N^2.
\label{H_b}
\end{eqnarray}
One recognizes $H_s$ as the Hamiltonian of an oscillator with frequency
\begin{eqnarray}
\omega_s=\sqrt{k_s/m_s},
\end{eqnarray}
and $H_b$ as the Hamiltonian  of a chain of $N$ atoms labeled $i=1\cdots N$ with boundary atoms $i=1$
and $i=N$ connected by springs to the walls. 

We shall refer to $H_s$ and $H_b$ as the Hamiltonians of the system and bath, respectively, but of course
they have such meaning only for $t<0$ while the bath's boundary atom $i=0$ is fixed. For $t>0$ the overall
system is described by Hamiltonian (\ref{H}) which includes the sum of $H_s$ and $H_b$,
but also the coupling terms involving the coordinate and momentum  of the boundary atom.

As mentioned above, we assume that for $t<0$ the system and bath are weakly coupled to external
thermal baths with temperatures $T_0$ and $T$, respectively.
As a result, the system acquires the canonical distribution
\begin{eqnarray}
\rho_s=Z^{-1}_s\,e^{-H_s/T_0},
\label{rho_s}
\end{eqnarray}
while the chain  acquires the  distribution
\begin{eqnarray}
\rho_b=Z_b^{-1}\,e^{-H_b/T}.
\label{rho_b}
\end{eqnarray}
Here and below we express temperature in the energy units so that Boltzmann's constant is unity,
$k_B=1$, Hamiltonians $H_s$ and $H_b$ are given by Eqs. (\ref{H_s}) and (\ref{H_b})
respectively, and $Z_s$, $Z_b$ are the partition functions of the  corresponding distributions.
At $t=0$ the connection to external baths and the constraint on the atom $i=0$ are removed,
and the overall system evolves as an isolated
mechanical system. Our goal is to find the internal energy of the system $U(t)$ 
\begin{eqnarray}
U(t)=\frac{1}{2m}\,\langle p^2(t)\rangle+\frac{k_s}{2}\,\langle [q(t)-q_0(t)]^2\rangle,
\end{eqnarray}
with an attention to the sign of the difference  $U(t)-U(0)$,
which determines the direction of the net heat exchange 
between the system and bath at a given time. Here and throughout the paper the angular brackets denote
the average over initial coordinates and momenta of the system and bath with the initial distribution
\begin{eqnarray}
\rho=\rho_s\, \rho_b.
\label{rho}
\end{eqnarray}
This distribution implies that the system and bath are initially prepared in uncorrelated canonical
states with  temperatures $T_0$ and $T$, respectively.

\section{Bath in terms of normal modes}

It is in many respects convenient and instructive to
make a canonical transformation of coordinates and momenta of the bath $\{q_i,p_i\}$ in order to 
diagonalize the bath Hamiltonian $H_b$, given by Eq. (\ref{H_b}), representing  it as a sum of
independent normal modes, see the right part of Fig. 1. Recall  that $H_b$  is the Hamiltonian of 
a uniform chain of $N$ atoms with terminal atoms $i=1$ and $i=N$ attached by springs $k$ to the walls.  
For such system, which is a linear version of the chain in the Fermi-Pasta-Ulam-Tsingou model,
the normal mode transformation 
\begin{eqnarray}
\{q_i,p_i\}\to \{P_j,Q_j\}, \qquad i,j=1\cdots N
\end{eqnarray}
is known to have the form
\begin{eqnarray}
q_i=\frac{1}{\sqrt{m}}\sum_{j=1}^N A_{ij} Q_j, 
\qquad
p_i=\sqrt{m}\sum_{j=1}^N A_{ij} P_j,\qquad
i=1\cdots N
\label{normal_modes_transformation}
\end{eqnarray}
with the transition matrix 
\begin{eqnarray}
A_{ij}=\sqrt{\frac{2}{N+1}} \sin \frac{\pi ij}{N+1}, \qquad i,j=1\cdots N
\end{eqnarray}
satisfying 
the orthogonality relation $\sum_{i=1}^N A_{ij}A_{ij'}=\delta_{jj'}$.
In terms of new coordinates $\{Q_j\}$ and momenta $\{P_j\}$ the Hamiltonian $H_b$  is diagonalized
into a sum of $N$ independent oscillators, or normal modes, with frequencies $\omega_j$,
\begin{eqnarray}
H_b=\frac{1}{2}\sum_{j=1}^N \Big\{ P_j^2+\omega_j^2 Q_j^2\Big\}, \qquad 
\omega_j=\omega_0\,\sin \frac{\pi j}{2(N+1)},
\label{H_b2}
\end{eqnarray}
where the characteristic frequency 
\begin{eqnarray}
\omega_0=2\sqrt{k/m}
\end{eqnarray}
has the meaning of the highest normal mode frequency 
in the infinite chain.

For $t>0$ (after the boundary atom is released) the Hamiltonian $H$ of the overall system is given by Eq. (\ref{H}).
It can be recomposed as
\begin{eqnarray}
H=H_0+H_b+H_c,
\label{H2}
\end{eqnarray}
where $H_0$ involves only variables of the system (isotope) and the boundary atom $i=0$ of the bath,
\begin{eqnarray}
H_0=\frac{p^2}{2m_s}+\frac{p_0^2}{2m}+\frac{k_s}{2}\,(q-q_0)^2+\frac{k}{2}\,q_0^2,
\end{eqnarray}
the bath's Hamiltonian $H_b$ is given by Eq. (\ref{H_b}) or Eq. (\ref{H_b2}), 
and $H_c$ describes the bilinear coupling of the boundary atom with the rest of the bath,
\begin{eqnarray}
H_c=-k\,q_0\,q_1.
\end{eqnarray}
Expressing $q_1$ in terms of normal modes with Eq. (\ref{normal_modes_transformation}), $H_c$ 
can be expressed as
\begin{eqnarray}
H_c=-q_0\sum_{j=1}^N c_jQ_j,
\label{H_c2}
\end{eqnarray}
with the coupling coefficients
\begin{eqnarray}
c_j=\frac{k}{\sqrt{m}}\,A_{1j}=\frac{k}{\sqrt{m}}\,\sqrt{\frac{2}{N+1}}\,\sin \frac{\pi j}{N+1}.
\label{c}
\end{eqnarray}
With bath variables expressed 
in terms of the normal modes, the overall system 
can be viewed as a two-atom cluster composed by the isotope and the boundary atom $i=0$,
the latter being 
bilinearly coupled by bounds of strength $c_j$ 
to $N$ independent oscillators with frequencies $\omega_j$. Such view, which makes the connection to the popular
Caldeira-Leggett model~\cite{Weiss}, is illustrated on  the right side  of Fig. 1.

\section{Langevin equation for the boundary atom}

We describe the overall system using  natural coordinates and momenta for the system $(q,\,p)$ and boundary atom
$(q_0,\,p_0)$, and normal mode coordinates and momenta
$\{Q_j,\,P_j\}$ for the bath. Such description is illustrated on the right part of Fig. 1 and corresponds to Hamiltonian
(\ref{H2}) with $H_b$ and $H_c$ in the normal mode representation given by Eqs. (\ref{H_b2}) and (\ref{H_c2}). 
Equations of motion have the form
\begin{eqnarray}
\dot p&=&-\frac{\partial H_0}{\partial q}=
-k_s(q-q_0),
\label{eom1}
\\
\dot p_0&=&
-\frac{\partial}{\partial q_0}(H_0+H_c)=
-k_s(q_0-q)-k\,q_0 +\sum_{j=1}^N c_j Q_j,
\label{eom2}\\
\dot P_j&=&
-\frac{\partial}{\partial Q_j}(H_b+H_c)=
-\omega_j^2\,Q_j+c_j\,q_0, \qquad j=1\cdots N.
\label{aux1}
\end{eqnarray}
Since $\dot Q_j=\partial H_b/\partial P_j=P_j$, the last equation (\ref{aux1}) can be written 
in terms of normal coordinates $Q_j$,
\begin{eqnarray}
\ddot Q_j=-\omega_j^2\,Q_j+c_j\, q_0, \qquad j=1\cdots N.
\label{eom3}
\end{eqnarray}
Solving Eqs. (\ref{eom3}) 
for $Q_j(t)$ and substituting the results into 
Eq. (\ref{eom2}), one can write the equation of motion of the boundary atom
in the form of the generalized Langevin equation,
\begin{eqnarray}
\dot p_0(t)&=&-k_s\,[q_0(t)-q(t)]
-\int_0^t K_0(t-t')\,p_0(t')\,dt' +\eta(t).
\label{le}
\end{eqnarray}
Except the term $-k(q_0-q)$, this equation is 
the familiar Langevin  equation for the  terminal atom in a semi-infinite harmonic chain, see e.g.~\cite{Weiss}.
In order to make the paper self-contained we provide 
details of the derivation of Eq. (\ref{le}) in Appendix A.

In  Eq. (\ref{le}), 
$\eta(t)$ is a fluctuating force
for which an explicit expression is available 
in the form of a 
linear function of initial coordinates and momenta of the bath, see Eq. (A7) in  Appendix A. 
With that expression, 
one can show  that the fluctuating force  is
zero-centered and stationary, 
\begin{eqnarray}
\langle \eta(t)\rangle=0, \qquad
\langle \eta(t')\,\eta(t'+t)\rangle=
\langle \eta(0)\,\eta(t)\rangle,
\end{eqnarray}
and related to 
the memory kernel $K_0(t)$ by the standard fluctuation-dissipation relations,
\begin{eqnarray}
\langle \eta(t)\,\eta(t')\rangle=m\, T \,K_0(t-t').
\label{fdt_0}
\end{eqnarray}
Here, as before, the angular brackets denote the averaging over the initial coordinates and momenta
of both the system and bath  
with the distribution $\rho=\rho_s\,\rho_b$. However, since $\eta(t)$ depends on bath variables only,
the average in above expressions is taken, in effect, with the bath distribution $\rho_b$ only.

As shown in Appendix A, the memory kernel $K_0(t)$ in the Langevin equation (\ref{le}) in the limit of
the infinite bath can be expressed in terms of the Bessel functions of the first kind $J_n(x)$ as 
\begin{eqnarray}
K_0(t)=\frac{\omega_0^2}{4}\,[J_0(\omega_0 t)+J_2(\omega_0 t)]=\omega_0\, \frac{J_1(\omega_0 t)}{2t},
\label{K_0}
\end{eqnarray}
where  the second expression is defined at $t=0$ by continuity.
One recognizes $K_0(s)$ as the kernel in the  generalized Langevin equation for a terminal
atom  of a semi-infinite harmonic chain~\cite{Weiss}. Note also that  expression (\ref{K_0})
is two times smaller than the kernel for a Langevin equation describing  
 a tagged atom   in the infinite homogeneous chain, see e.g.~\cite{Plyukhin}.
We shall need the Laplace transform of $K_0(t)$,
\begin{eqnarray}
\tilde K_0(s)=
\mathcal L\{K_0(t)\}=\int_0^\infty e^{-st} K_0(t)\,dt
\label{laplace}
\end{eqnarray}
which has the form
\begin{eqnarray}
\tilde K_0(s)=\frac{\omega_0^2/2}{s+\sqrt{s^2+\omega_0^2}}= \frac{1}{2}\Big(\sqrt{s^2+\omega_0^2}-s
\Big).
\label{K_0_laplace}
\end{eqnarray}

In our approach, the Langevin equation (\ref{le}) for the boundary atom plays an auxiliary role.
Our next goal in the next section will be  to integrate that equation  (in the Laplace domain) and,
substituting the result into the equation of motion (\ref{eom1}), to derive a Langevin equation for the system.

\section{Non-stationary Langevin equation for the system}

With bath degrees of freedom being integrated (see the previous section),
the system of relevant dynamical equations  is reduced to two equations, namely  the equation of
motion for the system  and the Langevin equation 
for the bath's boundary atom:
\begin{eqnarray}
\dot p(t)&=&
-k_s q(t)+k_s q_0(t),\label{eq1}\\
\dot p_0(t)&=&-k_s q_0(t)+k_s q(t)
-\int_0^t K_0(t-t')\,p_0(t')\,dt' +\eta(t).\label{eq2}
\end{eqnarray}
The initial conditions for the boundary atom, which is initially fixed, are 
\begin{eqnarray}
q_0(0)=p_0(0)=0,
\end{eqnarray}
while initial coordinate $q(0)$ and momentum $p(0)$ of the system
may be  arbitrary.  Later, we shall assume that $q(0),\,p(0)$ are drawn from a canonical
ensemble with the distribution $\rho_s$ given by Eq. (\ref{rho_s}).

Expressing coordinates in terms of momenta,
\begin{eqnarray}
q(t)=q(0)+\frac{1}{m_s}\,\int_0^t p(\tau)\, d\tau, \quad
q_0(t)=\frac{1}{m}\,\int_0^t p_0(\tau)\, d\tau,
\label{qp}
\end{eqnarray}
we can write Eqs.  (\ref{eq1}) and (\ref{eq2}) in the Laplace $s$-domain  as follows:
\begin{eqnarray}
s\,\tilde p(s)-p(0)&=& -k_s\,\frac{q(0)}{s}-\frac{k_s}{m_s}\,\frac{\tilde p(s)}{s}+\frac{k_s}{m}\,
\frac{\tilde p_0(s)}{s},\\
s\,\tilde p_0(s)&=&-\frac{k_s}{m}\,\frac{\tilde p_0(s)}{s}+k_s\,\frac{q(0)}{s}+
\frac{k_s}{m_s}\,\frac{\tilde p(s)}{s}-\tilde K_0(s)\,\tilde p_0(s)+\tilde\eta(s).
\end{eqnarray}
Here, the symbol tilde denotes the Laplace transforms defined in the standard way as in Eq. (\ref{laplace}),
and the Laplace variable $s$ should not be confused with with
the $s$ (``system") subscript of parameters $m_s$ and $k_s$.
Eliminating from the above equations $\tilde p_0$, one finds
\begin{eqnarray}
s\,\tilde p(s)-p(0)=-\tilde K(s)\,\tilde p(s)+\tilde \xi(s)-m_s\,q(0)\,\tilde K(s),
\label{p1} 
\end{eqnarray}
with 
\begin{eqnarray}
\tilde K(s)&=&
\frac{\alpha\,\beta\,\omega_0^2}{4}\,
\frac{s+\tilde K_0(s)}{s^2+s\,\tilde K_0(s)+\beta\,\omega_0^2/4},
\label{K}\\
\tilde \xi(s)&=&
\frac{\beta\,\omega_0^2/4}{s^2+s\, \tilde K_0(s)+\beta\,\omega_0^2/4}\,\,
\tilde\eta(s). 
\label{xi}
\end{eqnarray}
Note that in the right-hand side of Eq. (\ref{K}) the first factor has the meaning of
the square of the internal frequency $\omega_s$ of the system,
\begin{eqnarray}
\omega_s^2=\frac{k_s}{m_s}=\frac{\alpha\,\beta}{4}\,\omega_0^2,
\label{freq_s}
\end{eqnarray}
so we can write the expression for $\tilde K(s)$ a bit more compactly as
\begin{eqnarray}
\tilde K(s)=
\omega_s^2\,
\frac{s+\tilde K_0(s)}{s^2+s\,\tilde K_0(s)+\beta\,\omega_0^2/4}.
\label{K_comp}
\end{eqnarray}
In the time domain Eq. (\ref{p1}) has the form of the generalized Langevin equation for the system
\begin{eqnarray}
\dot p(t)=-\int_0^t K(t-\tau)\,p(\tau)\,d\tau +\xi(t)-m_s\,q(0)\,K(t),
\label{GLE}
\end{eqnarray}
where the dissipative memory kernel $K(t)$ and the fluctuating force $\xi(t)$ are defined
by their Laplace transforms (\ref{K}) and (\ref{xi}), respectively.

Equation (\ref{GLE})
plays the central role in our approach. Compared to a generalized Langevin equation of
the standard form~\cite{Weiss,Hanggi},  it has two special features. The first one is the presence
of the force $-m_s\,q(0)\, K(t)$ which depends on the initial position of the system and is
often referred to as the initial slip~\cite{Hanggi}. The presence in the Langevin equation of
additional terms depending on initial conditions appears to be a generic feature when the bath
is not in (constraint) equilibrium with 
the system~\cite{Hynes1,Hynes2,Bez}.
The second feature is that the fluctuating force $\xi(t)$ is not stationary. This can be seen
from Eq. (\ref{xi}), which shows that in the time domain $\xi(t)$ is a convolution  of
stationary noise $\eta(t)$ and thus in general is not stationary itself. 
The non-stationarity of $\xi(t)$ will be confirmed below, in particular,
by the explicit evaluation of the second moment
$\langle \xi^2(t)\rangle$, which will be shown to be  time-dependent.
In contrast, the noise $\eta(t)$ in the Langevin equation (\ref{le}) for the boundary
atom is stationary and, according to Eq. (\ref{fdt_0}), has a time-independent second 
moment $\langle \eta^2(t)\rangle=m\,T\, K_0(0)$.
Physically,  the non-stationarity of $\xi(t)$ is to be expected because the force on the system
is exerted by the boundary atom, which is not in an equilibrium or stationary state after being released at $t=0$.

For a non-stationary noise $\xi(t)$, the correlation $\langle \xi(t)\, \xi(t')\rangle$
is not a function of the time difference $t-t'$ only. Clearly, in that case the 
standard fluctuation-dissipation relation like Eq. (\ref{fdt_0}) cannot be valid,
\begin{eqnarray}
\langle \xi(t)\, \xi(t')\rangle
\ne m_s T K(t-t').
\end{eqnarray}
We shall address the derivation  of an adequate  relation between $\xi(t)$ and $K(t)$ in the next section.
Meanwhile, let us  discuss the properties of those functions separately.

Using Eqs. (\ref{K}) and (\ref{K_0_laplace}), the Laplace transform of the memory kernel $\tilde K(s)$
can be brought to the following more explicit form:
\begin{eqnarray}
\tilde K(s)=\frac{\alpha\,\beta\,\omega_0^2}{4}\,
\frac{s+\sqrt{s^2+\omega_0^2}}{s^2+s\,\sqrt{s^2+\omega_0^2}+\beta\,\omega_0^2/2}.
\end{eqnarray}
By factorizing the denominator
\begin{eqnarray}
s^2+s\, \sqrt{s^2+\omega_0^2} +\beta\,\omega_0^2/2=\frac{1}{2}\,\left(\sqrt{s^2+\omega_0^2}+s\right)\,
\left(\beta\,\sqrt{s^2+\omega_0^2}+(2-\beta)\,s
\right),
\end{eqnarray}
the expression is further 
simplified to 
\begin{eqnarray}
\tilde K(s)=\frac{\alpha\,\beta\,\omega_0^2/2}{\beta\,\sqrt{s^2+\omega_0^2}+(2-\beta)\, s}.
\label{KK}
\end{eqnarray}
In the time domain, the kernel $K(t)$ is available in the closed form only 
for $\beta=1$ and $\beta=2$. 
For $\beta=1$, Eq. (\ref{KK}) reads as
\begin{eqnarray}
\tilde K(s)=\frac{\alpha\,\omega_0^2/2}{\sqrt{s^2+\omega_0^2}+s}=\alpha\, \tilde K_0(s), \qquad \beta=1.
\end{eqnarray}
Thus, for $\beta=1$ the kernel $K(t)$ in the Langevin equation for the system differs from
that for the boundary atom $K_0(t)$
just by the factor $\alpha$,
\begin{eqnarray}
K(t)=\alpha\, K_0(t)=\frac{\alpha\,\omega_0^2}{4}\,
[J_0(\omega_0 t)+J_2(\omega_0 t)]=
\alpha\,\omega_0\,\frac{J_1(\omega_0 t)}{2t}, \qquad \beta=1.
\label{K1}
\end{eqnarray}
For $\beta=2$, Eq. (\ref{KK}) takes the form
\begin{eqnarray}
\tilde K(s)=\frac{\alpha\,\omega_0^2/2}{\sqrt{s^2+\omega_0^2}}, \qquad \beta=2.
\label{K2}
\end{eqnarray}
In the time domain this corresponds to
\begin{eqnarray}
K(t)=\frac{\alpha\,\omega_0^2}{2}
\,J_0(\omega t), \qquad \beta=2.
\label{K22}
\end{eqnarray}
In both cases $\beta=1$ and $\beta=2$ the kernel $K(t)$ is a decaying oscillatory function,
with the oscillation amplitude decaying as $t^{-3/2}$ and $t^{-1/2}$, respectively.
Such asymptotic behavior can be viewed as an  example of a general trend that the stronger
the system is coupled to the bath, the slower is the decay of relevant correlation functions.
The connection between  the kernel $K(t)$ and the correlation function of the noise will
be discussed in the next section.

An important property of $K(t)$ is its
initial value $K(0)$. For any values of $\alpha$ and $\beta$ we find from Eq. (\ref{KK})
\begin{eqnarray}
K(0)=\lim_{s\to\infty} s\,\tilde K(s)=
\frac{\alpha\,\beta}{4}\, \omega_0^2=\omega_s^2,
\label{K_initial}
\end{eqnarray}
where, recall,  $\omega_s=\sqrt{k_s/m_s}$ is the internal frequency of the system.

Now consider properties of the fluctuating force $\xi(t)$. As  follows from Eq. (\ref{xi}), in the time
domain $\xi(t)$  is given by the convolution 
\begin{eqnarray}
\xi(t)=\int_0^t L(t-\tau)\,\eta(\tau)\,d\tau
\label{xi2}
\end{eqnarray}
of the stationary noise $\eta(t)$ 
and the kernel $L(t)$ 
with the Laplace transform 
\begin{eqnarray}
\tilde L(s)=\frac{\beta\,\omega_0^2/4}{s^2+s\,\tilde K_0(s)+\beta\,\omega_0^2/4}.
\label{L}
\end{eqnarray}
As follows from Eq. (\ref{xi2}),
since the noise $\eta(t)$ is zero-centered then so is $\xi(t)$,
\begin{eqnarray}
\langle \xi(t)\rangle=0.
\end{eqnarray}
In order to evaluate time correlations and moments of $\xi(t)$
we need to discuss  properties of the kernel $L(t)$ and its connections with 
kernels $K_0(t)$ and $K(t)$ in the Langevin equations for the boundary atom and system, respectively.

Substituting  expression (\ref{K_0}) for $\tilde K_0(s)$
into Eq. (\ref{L})
yields $\tilde L(s)$ as an explicit function of $s$,
\begin{eqnarray}
\tilde L(s)=\frac{\beta\,\omega_0^2/2}{s^2+s\sqrt{s^2+\omega_0^2}+\beta\,\omega_0^2/2}.
\label{L2}
\end{eqnarray}
From here  we find that
the initial value of $L(t)$ is zero,
\begin{eqnarray}
L(0)=\lim_{s\to\infty} s\,\tilde L(s)=0.
\label{L_initial}
\end{eqnarray}

Next, using expressions (\ref{L}) and (\ref{K_comp}) for $\tilde L(s)$ and $\tilde K(s)$ one finds
that in the time domain $L(t)$ is given by a negative derivative of the dissipative kernel $K(t)$, 
\begin{eqnarray}
L(t)=-\omega_s^{-2}\,\dot  K(t).
\label{property_2}
\end{eqnarray}
Indeed, multiplying Eq. (\ref{K_comp}) by $s$ and then  adding and subtracting $\beta\,\omega_0^2/4$
in the numerator, one gets
\begin{eqnarray}
s\,\tilde K(s)=\omega_s^2\,
[1-\tilde L(s)].
\label{chi_aux}
\end{eqnarray}
Recalling that $\omega_s^2$ is the initial value 
of the kernel $K(t)$, see Eq. (\ref{K_initial}), the above relation can be written as
\begin{eqnarray}
\tilde L(s)=-\omega_s^{-2}\,[s\,\tilde K(s)-K(0)].
\label{chi3}
\end{eqnarray}
In the time domain this gives Eq. (\ref{property_2}).

Another useful property is the relation between the kernels $L(t)$, $K(t)$, and $K_0(t)$ in the Laplace domain,
\begin{eqnarray}
\tilde K(s)=\alpha\,s\,\tilde L(s)+\alpha\,\tilde L(s)\, \tilde K_0(s).
\label{property_1}
\end{eqnarray}
This follows directly from
expressions (\ref{K}) and (\ref{L})
for $\tilde K(s)$ and $\tilde L(s)$.
Since $L(0)=0$, in the above expression $s\tilde L(s)$ is the transform of $\dot L(t)$.
Therefore, in the time domain relation (\ref{property_1}) reads
\begin{eqnarray}
K(t)=\alpha\,\dot L(t)+\alpha\,\int_0^{t}d\tau \,L(\tau)\,K_0(t-\tau).
\label{property_11}
\end{eqnarray}

To finish this section, let us use the above relations to evaluate the second moment of the noise $\xi(t)$,
\begin{eqnarray}
\langle \xi^2(t)\rangle=
\int_0^t d\tau_1\, L(\tau_1)
\int_0^t d\tau_2\, L(\tau_2)\,
\langle \eta(t-\tau_1)\,\eta(t-\tau_2)\rangle.
\end{eqnarray}
Since the  noise $\eta(t)$ is stationary and satisfies the fluctuation-dissipation relation (\ref{fdt_0}), 
the above expression
takes the form
\begin{eqnarray}
\langle \xi^2(t)\rangle=m\,T
\int_0^t d\tau_1\, L(\tau_1)
\int_0^t d\tau_2\, L(\tau_2)\,K_0(\tau_2-\tau_1),
\end{eqnarray}
or
\begin{eqnarray}
\langle \xi^2(t)\rangle=2m\,T
\int_0^t d\tau_1 L(\tau_1)
\int_0^{\tau_1}d\tau_2\, L(\tau_2)\,K_0(\tau_1-\tau_2).
\label{moment2}
\end{eqnarray}
Here, the inner integral is the convolution of $L(t)$ and $K_0(t)$,
which can be found from Eq. (\ref{property_11}), 
\begin{eqnarray}
\langle \xi^2(t)\rangle=\frac{2\,m\,T}{\alpha}\int_0^t d\tau\, L(\tau)K(\tau)-2\,m\,T\,\int_0^t d\tau\, L(\tau)\,\dot L(\tau).
\label{B7}
\end{eqnarray}
Next we use property (\ref{property_2}) to get
\begin{eqnarray}
\langle \xi^2(t)\rangle=-\frac{2\,m\,T}{\alpha\,
\omega_s^2}\int_0^t d\tau\, K(\tau)\dot K(\tau)-2\,m\,T\,\int_0^t d\tau L(\tau)\dot L(\tau).
\end{eqnarray}
The integration yields
\begin{eqnarray}
\langle \xi^2(t)\rangle=-\frac{m\,T}{\alpha\,\omega_s^2}\,[K^2(t)-K^2(0)]-m\,T\,L^2(t),
\label{B9}
\end{eqnarray}
where we recall that  $L(0)=0$. Substituting here $L(t)$ in the form (\ref{property_2}) and 
$K(0)=\omega_s^2$
finally yields
\begin{eqnarray}
\langle \xi^2(t)\rangle=m_s\,\omega_s^2\, T-\frac{m_s\,T}{\omega_s^2}\,K^2(t)-
\frac{m\,T}{\omega_s^4}\,[\dot K(t)]^2.
\label{xi_moment}
\end{eqnarray}
This expression shows explicitly and quantifies the non-stationarity of  the noise $\xi(t)$ and
its connection to the dissipative kernel $K(t)$. In the next section we shall be able to derive
this expression in a more general way from the fluctuation-dissipation  relation for
the correlation $\langle\xi(t)\,\xi(t')\rangle$.

Note  also that although for $t>0$ the function  $\xi(t)=\int_0^t L(t-\tau)\,\eta(\tau)\,d\tau$ 
fluctuates, at $t=0$ it takes a pre-determined zero value, $\xi(0)=0$. 
This is consistent with Eq. (\ref{xi_moment}), which gives $\langle \xi^2(0)\rangle=0$,
taking into account that $K(0)=\omega_s^2$ and
$\dot K(0)=0$.

\section{Nonstationary fluctuation-dissipation relation}

In this section we shall find a (fluctuation-dissipation) relation between  the correlation function
of the non-stationary noise $\xi(t)$ and the dissipative kernel $K(t)$ in  the Langevin equation
(\ref{GLE}) for the system. Such a  relation is of interest of its own, but we shall also use it
in sections to follow to evaluate the system's internal energy as a function of time.

Recall that $\xi(t)$ is given by the convolution integral
$\xi(t)=\int_0^t L(t-\tau)\, \eta(\tau)\, 
d\tau$, 
where $\eta(t)$ is the noise in the Langevin equation (\ref{le}) for the boundary atom. The noise $\eta(t)$
is stationary and satisfies the standard fluctuation-dissipation relation (\ref{fdt_0}), 
 $\langle \eta(t)\,\eta(t')\rangle=mT\,K_0(t-t')$.
 Then the two-time correlation function of $\xi(t)$ is
\begin{eqnarray}
\langle \xi(t_1)\,\xi(t_2)\rangle=m\,T\int_0^{t_1} d\tau_1\int_0^{t_2} d\tau_2\, L(t_1-\tau_1)\,L(t_2-\tau_2)\,K_0(\tau_2-\tau_1).
\label{corr1}
\end{eqnarray}
This expression has the form of the double convolution
\begin{eqnarray}
(f*\!*\,g)(t_1,t_2)\equiv
\int_0^{t_1} d\tau_1\int_0^{t_2} d\tau_2\, f(t_1-\tau_1,\,t_2-\tau_2)\,\,g(\tau_1,\tau_2)
\label{conv1}
\end{eqnarray}
of the two-variable functions
\begin{eqnarray}
f(t_1,t_2)=L(t_1)\,L(t_2), \qquad g(t_1,t_2)=m\,T\,K_0(t_2-t_1).
\end{eqnarray}
A convenient mathematical tool to handle expressions with double convolutions is  the double
Laplace transform of a two-variable function $f(t_1,t_2)$,
\begin{eqnarray}
\mathcal L_2\{f(t_1,t_2)\}\equiv\int_0^\infty \!\!\!\! dt_1\, e^{-s_1 t_1}\!\!\int_0^\infty \!\!\!\! dt_2\,e^{-s_2 t_2} \,f(t_1,t_2).
\end{eqnarray}
The convolution theorem for the double Laplace transforms reads as
\begin{eqnarray}
\mathcal L_2\{f\!*\!*\,g\}=\mathcal{L}_2\{f\}\,\mathcal{L}_2\{g\},
\label{conv_th}
\end{eqnarray}
see, e.g., Ref. \cite{Debnath}.
Applying the theorem to the double convolution (\ref{corr1}) yields
\begin{eqnarray}
\mathcal L_2\{\langle \xi(t_1)\,\xi(t_2)\rangle\}=
m\,T\,\mathcal L_2\{L(t_1)\,L(t_2)\}\,\,\mathcal L_2\{K_0(t_2-t_1)\}.
\label{corr2}
\end{eqnarray}
It is clear that 
\begin{eqnarray}
\mathcal L_2\{L(t_1)\,L(t_2)\}=\mathcal L\{L(t_1)\}\,\mathcal L\{L(t_2)\}=\tilde L(s_1)\,\tilde L(s_2),
\label{L2_prop1}
\end{eqnarray}
where  $\mathcal L$ and the tilde denote, as in the previous sections,   the Laplace transform of a single  variable function,
$\mathcal L\{f(t)\}=\tilde f(s)=\int_0^\infty e^{-st} f(t)\,dt$.
Therefore, Eq. (\ref{corr2}) takes the form
\begin{eqnarray}
\mathcal L_2\{\langle \xi(t_1)\,\xi(t_2)\rangle\}=
m\,T\,\tilde L(s_1)\,\tilde L(s_2)\,\,\mathcal L_2\{K_0(t_2-t_1)\}.
\label{corr3}
\end{eqnarray}
Next we use the following property of the double Laplace transform for an even function~\cite{Debnath}:
\begin{eqnarray}
\mathcal L_2\{f(t_2-t_1)\}=\frac{1}{s_1+s_2}\,[\tilde f(s_1)+\tilde f(s_2)], \quad \mbox{if} \quad f(t)=f(-t).
\label{L2_prop2}
\end{eqnarray}
According to Eq. (\ref{fdt_0}),  the kernel $K_0(t)$ is proportional to the correlation function of a stationary noise $\eta(t)$
and therefore is an even function. Then applying Eq. (\ref{L2_prop2}) we get
\begin{eqnarray}
\mathcal L_2\{ K_0(t_2-t_1)\}=\frac{1}{s_1+s_2}\,\{\tilde K_0(s_1)+\tilde K_0(s_2)\},
\end{eqnarray}
and Eq. (\ref{corr3}) takes the form
\begin{eqnarray}
\mathcal L_2\{\langle \xi(t_1)\,\xi(t_2)\rangle\}=
m\,T\,\frac{\tilde L(s_1)\,\tilde L(s_2)}{s_1+s_2}\,\,[\tilde K_0(s_1)+\tilde K_0(s_2)].
\label{corr4}
\end{eqnarray}
The next step is to use relation (\ref{property_1}), which we can write as
\begin{eqnarray}
\tilde L(s)\,\tilde K_0(s)=\frac{1}{\alpha}\,\tilde K(s)-s\,\tilde L(s).
\label{property_x}
\end{eqnarray}
From Eqs. (\ref{corr4}) and (\ref{property_x}) one gets
\begin{eqnarray}
\mathcal L_2\{\langle \xi(t_1)\,\xi(t_2)\rangle\}=
\frac{m\,T}{\alpha}\,\frac{\tilde L(s_1)\,\tilde K(s_2)+\tilde L(s_2)\tilde K(s_1)}{s_1+s_2}
-m\,T\,\tilde L(s_1)\,\tilde L(s_2).
\label{corr5}
\end{eqnarray}
The inverse transform $\mathcal L_2^{-1}$ of this expression is
\begin{eqnarray}
\langle \xi(t_1)\,\xi(t_2)\rangle=\frac{m\,T}{\alpha}\, (f*\!*\,g)(t_1,t_2)-m\,T\,L(t_1)\,L(t_2),
\label{corr6}
\end{eqnarray}
where the double convolution $(f\!*\!*\,g)$ involves the functions
\begin{eqnarray}
f(t_1,t_2)&=&\mathcal{L}_2^{-1}\left\{\frac{1}{s_1+s_2}\right\}=
\delta(t_2-t_1),\\ 
g(t_1,t_2)&=&\mathcal L_2^{-1}\Bigl\{ \tilde L(s_1)\,\tilde K(s_2)+\tilde L(s_2)\tilde K(s_1)\Bigr\}=L(t_1)\,K(t_2)+L(t_2)\,K(t_1).
\label{g}
\end{eqnarray}
One can verify that the double convolution of $f(t_1,t_2)=\delta(t_2-t_1)$ and an arbitrary function $g(t_1,t_2)$ is
\begin{eqnarray}
(f*\!*\,g)(t_1,t_2)=\int\limits_0^{min(t_1,\, t_2)} \!\!\!\!\!\!g(t_1-\tau',t_2-\tau')\,d\tau'.
\label{aux2}
\end{eqnarray}
In our case, the function  $g(t_1,t_2)$ is given by Eq. (\ref{g}) and has the symmetry property $g(t_1,t_2)=g(t_2,t_1)$,
which allows a further simplification. Making in Eq. (\ref{aux2}) the substitutions $\tau=t_1-\tau'$ for $t_2>t_1$
and $\tau=t_2-\tau'$ for $t_1>t_2$, in other words, $\tau=min(t_1,\,t_2)-\tau'$, one gets
\begin{eqnarray}
(f*\!*\,g)(t_1,t_2)=\int\limits_0^{min(t_1,\, t_2)} \!\!\!\!\!\!g\bigl(|t_2-t_1|+\tau,\tau\bigr)\,d\tau,
\end{eqnarray}
or more explicitly
\begin{eqnarray}
(f*\!*\,g)(t_1,t_2)=\int\limits_0^{min(t_1, t_2)} \!\!\!\!\!\!\Bigl\{
L\bigl(|t_2-t_1|+\tau\bigr)\,K(\tau)+
L(\tau)\,K\bigl(|t_2-t_1|+\tau\bigr)\Bigr\}\,d\tau.
\label{conv3}
\end{eqnarray}
Next, recall that the kernels $K(t)$ and $L(t)$ are connected by relation (\ref{property_2}),
\begin{eqnarray}
 L(t)=
-\omega_s^{-2}\,\dot K(t), 
\label{aux5}
\end{eqnarray}
where $\omega_s=\sqrt{k_s/m_s}$ is the internal frequency of the system.
Combining Eqs. (\ref{corr6}), (\ref{conv3}) and (\ref{aux5}) yields
\begin{eqnarray}
\langle \xi(t_1)\,\xi(t_2)\rangle
=-\frac{m\,T}{\alpha\,\omega_s^2}\,
\int\limits_0^{min(t_1,t_2)} \!\!\!\!\!\!\Bigl\{
\dot K\bigl(|t_2-t_1|+\tau\bigr)\,K(\tau)+
\dot K(\tau)\,K\bigl(|t_2-t_1|+\tau\bigr)\Bigr\}\,d\tau
-\frac{m\,T}{\omega_s^4}\,\dot K(t_1)\,\dot K(t_2).
\end{eqnarray}
Noticing that here the integrand is the total derivative $\frac{d}{d\tau}\,[K\bigl(|t_2-t_1|+\tau\bigr)\,K(\tau)]$
and recalling that $K(0)=\omega_s^2$, we finally obtain
\begin{eqnarray}
\langle \xi(t_1)\,\xi(t_2)\rangle
=m_s\,T\,K\bigl(|t_2-t_1|\bigr)
-\frac{m_s\,T}{\omega_s^2}
\,K(t_1)K(t_2)-\frac{m\,T}{\omega_s^4}\,\dot K(t_1)\,\dot K(t_2).
\label{FDR}
\end{eqnarray}
This is the fluctuation-dissipation relation for the present model. For $t_1=t_2$ it
gives expression (\ref{xi_moment}) for the second moment of the fluctuation force $\langle  \xi^2(t)\rangle$,
which we derived previously by another method.

The last two terms in Eq. (\ref{FDR}) are not functions of the time difference and thus reflect the
non-stationarity of the noise $\xi(t)$.  We observe that for the present model the dependence of the
non-stationary terms on $t_1$ and $t_2$ is  simply factorized.  
If the kernel $K(t)$ and its first derivative vanish at long times, the non-stationary terms in Eq. (\ref{FDR})
vanish faster than the stationary one, and the noise  $\xi(t)$ becomes stationary at asymptotically long times.

In the next two sections we shall exploit the fluctuation-dissipation relation (\ref{FDR})
to evaluate the internal energy 
of the system as a function of time.
For that application, the non-stationary terms in Eq. (\ref{FDR}) are essential and cannot be neglected
even if they are relatively small at long times.
The reader not interested in the mathematical aspects of the evaluation may skip the next two sections
and go directly to Section IX where the results are summarized and discussed.

\section{Kinetic energy}

In this section we evaluate the average kinetic energy of the system $E(t)=\langle p^2(t)\rangle/2m_s$,
solving the Langevin equation (\ref{GLE}),
\begin{eqnarray}
\dot p(t)=-\int_0^t K(t-\tau)\,p(\tau)\,d\tau +\xi(t)-m_s\,q(0)\,K(t).
\label{GLE2}
\end{eqnarray}
The evaluation of the second moment of a targeted 
stochastic variables, in our case $\langle p^2(t)\rangle$,  from a generalized Langevin
equation is a straightforward exercise provided 
the noise is  stationary and the fluctuation-dissipation relation has the standard form,
see e.g.~\cite{Plyukhin}. For the present model the noise $\xi(t)$ is non-stationary, and more elaboration is needed.

 The solution of the Langevin equation (\ref{GLE2}) in the Laplace domain reads
 \begin{eqnarray}
 \tilde p(s)=p(0)\,\tilde R(s)-m_s\,q(0)\,\tilde K(s)\,\tilde R(s)+\tilde \xi(s)\,\tilde R(s),
 \label{solution1}
 \end{eqnarray}
 where 
 \begin{eqnarray}
 \tilde R(s)=\frac{1}{s+\tilde K(s)}.
 \label{resolvent1}
 \end{eqnarray}
 We shall call the function $R(t)$ the resolvent. 
 It is also often called the relaxation function. 
 As follows from Eq. (\ref{solution1}), the physical meaning of the resolvent $R(t)$ is that it gives
 a solution  $p(t)$ for the specific initial condition when $p(0)=1$, $q(0)=0$, while  atoms of
 the bath are initially at rest in equilibrium positions,
 $p_i(0)=q_i(0)=0$ for $i=0\cdots N$ (in that case $\xi(t)=0$ at any time $t>0$, see appendix A).
 As will be discussed in the following sections, the resolvent
 may show either decaying  or oscillating behavior  at long times depending on specific values of
 parameters $\alpha$ and $\beta$. In this section we focus on general relations and make
 no assumptions about asymptotic properties of the resolvent at long times.

 Writing Eq. (\ref{resolvent1}) as $s\tilde R(s)-1=-\tilde K(s)\,\tilde R(s)$, one notices that in the
 time domain the function $R(t)$ satisfies the following initial value problem:
 \begin{eqnarray}
 \dot R(t)=-\int_0^t K(t-\tau)\,R(\tau)\,d\tau, \qquad R(0)=\lim_{s\to\infty}s\,\tilde R(s)=1.
 \label{resolvent2}
 \end{eqnarray}
 Here the initial condition follows from Eq. (\ref{resolvent1}) and expression (\ref{KK}) 
 for the kernel $\tilde K(s)$, 
 which  shows that 
 $\lim_{s\to\infty} \tilde K(s)=0$. 
  As follows  from Eq. (\ref{resolvent2}),  the Laplace transform  
 and initial value of the resolvent's first derivative
 are 
 \begin{eqnarray}
 \mathcal{L}\{\dot R(t)\}=-\tilde K(s)\,\tilde R(s),\qquad \dot R(0)=0.
 \label{dotT}
 \end{eqnarray}
 We shall also need the Laplace transform and initial value of the resolvent's second derivative.
 Since $\dot R(0)=0$ we get
 \begin{eqnarray}
 \mathcal{L}\{\ddot R(t)\}=s\,\mathcal{L}\{\dot R(t)\}=-s\,\tilde K(s)\,\tilde R(s),\qquad
\ddot R(0)=-\lim_{s\to\infty}s^2\tilde K(s)\tilde R(s)=-\omega_s^2.
\label{ddotR}
\end{eqnarray}
The latter relation follows from the initial value theorem and the  
asymptotic behavior of the kernel
for large $s$,
\begin{eqnarray}
\tilde K(s)\sim \frac{\alpha\,\beta\,\omega_0^2}{4\,s}=\frac{\omega_s^2}{s},\qquad s\to\infty,
\end{eqnarray}
see Eq. (\ref{KK}). Taking into account Eq. (\ref{dotT}), one observes that expression (\ref{solution1})
in the time domain  reads
 \begin{eqnarray}
 p(t)=p(0)\,R(t)+m_s\,q(0)\,\dot R(t)+\int_0^t R(t-t')\,\xi(t')\, dt'.
 \label{sol2}
 \end{eqnarray}
 For the last term here
 let us introduce a temporary notation
 \begin{eqnarray}
 p_0(t)\equiv\int_0^t R(t-t')\,\xi(t')\, dt'.
 \label{p_0}
 \end{eqnarray}
 The function $p_0(t)$ gives the system's momentum for initial conditions  with $q(0)=p(0)=0$.
 Since the noise $\xi(t)$ is zero-centered, the first moment of $p_0(t)$ vanishes, $\langle p_0(t)\rangle=0$.
 Then 
 squaring Eq. (\ref{sol2}) and taking the average with the distribution (\ref{rho}) we obtain
 \begin{eqnarray}
 \langle p^2(t)\rangle=
 \langle p^2\rangle\, R^2(t)+m_s^2\,\langle q^2\rangle [\dot R(t)]^2+2\,m_s\, \langle p\,q\rangle \,R(t)\dot R(t)+\langle p_0^2(t)\rangle.
 \label{qq1}
 \end{eqnarray}
 Here $q=q(0)$ and $p=p(0)$ are initial values of the system's variables. Their moments in Eq. (\ref{qq1}) are calculated, in effect, 
 with the distribution $\rho_s$ given by  Eq. (\ref{rho_s}),
 \begin{eqnarray}
 \langle p^2\rangle=m_s\,T_0, \qquad
 \langle q^2\rangle=\frac{1}{k_s}\,T_0=\frac{1}{m_s\,\omega_s^2}\,T_0,\qquad \langle q\,p\rangle=0,
 \label{moments}
 \end{eqnarray}
 where $T_0$ is the initial temperature of the system, 
 then 
 \begin{eqnarray}
 \langle p^2(t)\rangle=
 m_s\,T_0\,R^2(t)+
 \frac{m_s\,T_0}{\omega_s^2}\,[\dot R(t)]^2+\langle p_0^2(t)\rangle.
 \label{p2}
 \end{eqnarray}

 As the next step we need to work out the last term in the above expression,
 \begin{eqnarray}
 \langle p_0^2(t)\rangle=\int_0^t d\tau_1\, R(t-\tau_1)
 \int_0^t d\tau_2\, R(t-\tau_2) \,\langle \xi(\tau_1)\,\xi(\tau_2)\rangle.
 \label{p_02av}
 \end{eqnarray}
 Using the fluctuation-dissipation relation (\ref{FDR}) we get
  \begin{eqnarray}
 \langle p_0^2(t)\rangle
 =m_s\,T\,\int_0^t d\tau_1\, R(\tau_1)
 \int_0^t d\tau_2\, R(\tau_2) \,K\bigl(
 |\tau_2-\tau_1|\bigr)
 -\frac{m_s\,T}{\omega_s^2}
 \,\Bigl[(R*K)(t)\Bigr]^2
 -\frac{m\,T}{\omega_s^4}\,\Bigl[(R*\dot K)(t)\Bigr]^2.
 \end{eqnarray}
 Here we use the notation $(f*g)(t)$ for the convolution $\int_0^t f(t-\tau) g(\tau) d\tau$. 
 To proceed, let us denote the three terms in the right-hand side of the above expression as $A_i(t)$, 
 \begin{eqnarray}
 \langle p_0^2(t)\rangle
 =A_1(t)+A_2(t)+A_3(t),
 \label{sumofA}
 \end{eqnarray}
 and evaluate each term  separately.

 The first term $A_1$ can be worked out with the standard trick of replacing the integral
 over the square $(0,t)\times(0,t)$ by the  two times integral over a triangle,
 \begin{eqnarray}
 A_1(t)=m_s\,T\int_0^t d\tau_1\, R(\tau_1)
 \int_0^t d\tau_2\, R(\tau_2) \,K\bigl(
 |\tau_2-\tau_1|\bigr)=2m_s\,T\int_0^t d\tau_1\, R(\tau_1)
 \int_0^{\tau_1} d\tau_2\, R(\tau_2) \,K(
 \tau_1-\tau_2).
 \end{eqnarray}
 Here the inner integral, according to Eq. (\ref{resolvent2}), equals
 $-\dot R(\tau_1)$, then
 \begin{eqnarray}
A_1(t)=-2m_s\,T\int_0^t d\tau\, R(\tau)\,\dot R(\tau)=m_s\,T\bigl[1-R^2(t)\bigl].
 \end{eqnarray}
The second term, again due to Eq. (\ref{resolvent2}), is 
\begin{eqnarray}
A_2(t)=-
\frac{m_s\,T}{\omega_s^2}
 \,\Bigl[(R*K)(t)\Bigr]^2=
 -\frac{m_s\,T}{\omega_s^2}\,\bigl[\dot R(t)]^2.
\end{eqnarray}
In order to evaluate the third term 
 \begin{eqnarray}
A_3(t)=-
 \frac{m\,T}{\omega_s^4}\,\Bigl[(R*\dot K)(t)\Bigr]^2
 \end{eqnarray}
we use the relation
\begin{eqnarray}
(R*\dot K)(t)=
-\ddot R(t)-K(0)\,R(t),
\label{aux55}
\end{eqnarray}
which can be obtained by differentiating Eq. (\ref{resolvent2}), or by evaluating the convolution $R*\dot K$
in the Laplace domain using Eq. (\ref{ddotR}).
Recalling that  $K(0)=\omega_s^2$, one finds
 \begin{eqnarray}
A_3(t)=
 -m\,T\,\left[R(t)+
 \omega_s^{-2}\,\ddot R(t)\right]^2.
 \end{eqnarray}
Combining the above results according to Eq. (\ref{sumofA}), we obtain
\begin{eqnarray}
\langle p_0^2(t)\rangle=m_s T\,\bigl[1-R^2(t)\bigl]
 -\frac{m_s\,T}{\omega_s^2}\,\bigl[\dot R(t)]^2
-m\,T\,\left[R(t)+\omega_s^{-2}\,\ddot R(t)\right]^2.
 \end{eqnarray}
Substituting this into Eq. (\ref{p2}) yields 
\begin{eqnarray}
\langle p^2(t)\rangle=m_s\,T+m_s (T_0-T)\,\Bigl\{R^2(t)+
\omega_s^{-2}\,\dot R(t)^2
\Bigr\}
-m\,T\,\left\{R(t)+\omega_s^{-2}\,\ddot R(t)\right\}^2.
\label{P2_result}
 \end{eqnarray}
Then for  the system's average kinetic energy $E=\langle p^2\rangle/2m_s$ we finally get the following expression
\begin{eqnarray}
E(t)=\frac{T}{2}+\frac{T_0-T}{2}\,\left\{
R^2(t)+
\omega_s^{-2}\,\dot R(t)^2
\right\}
-\frac{\alpha\,T}{2}\,\left\{R(t)+\omega_s^{-2}\,\ddot R(t)\right\}^2.
\label{E}
 \end{eqnarray}

Since $R(0)=1$, $\dot R(0)=0$, and $\ddot R(0)=-\omega_s^2$, the above expression for $t=0$ gives $E(0)=T_0/2$,
which is the correct  equilibrium value for the given setup. The behavior of $E(t)$ at long times is governed by
asymptotic properties of the resolvent and its derivatives. For an  ergodic
system $R(t), \dot R(t), \ddot R(t)\to  0$ at  long times. Then Eq. (\ref{E}) describes, in accordance
with the equipartition theorem,  relaxation to the equilibrium value at the bath temperature $T$, 
$E(t)\to T/2$, while 
the last two terms in Eqs. (\ref{P2_result}) and (\ref{E}) describe the transient. 
Because of the last term, the transient is not identically zero even if $T=T_0$.
From the point of view of macroscopic thermodynamics this is an anomaly contradicting the zeroth law,
but microscopically this is a result to anticipate since the initial distribution (\ref{rho}) 
does not involve the system-bath interaction and 
is not the equilibrium distribution for the overall system even when $T_0=T$.


\section{Potential energy}
According to the equation of motion for the system (\ref{eom1}), $q(t)-q_0(t)=-\dot p(t)/k_s$. Then the average  
potential energy of the system can be written as
\begin{eqnarray}
V(t)=\frac{k_s}{2}\,\langle [q(t)-q_0(t)]^2\rangle=
\frac{1}{2\,k_s}\,\langle \dot p(t)^2\rangle.
\label{V0}
\end{eqnarray}
Differentiating solution (\ref{sol2}) of the Langevin equation we get
\begin{eqnarray}
 \dot p(t)=p(0)\,\dot R(t)+m_s\,q(0)\,\ddot R(t)+\xi(t)+\varphi(t), 
 \label{sol3}
 \end{eqnarray}
 where the last term is a new fluctuating force defined as
\begin{eqnarray}
 \varphi(t)=\int_0^t \dot R(t-t')\,\xi(t')\, dt'.
 \label{phi}
 \end{eqnarray}
Both fluctuating forces $\xi(t)$ and $\varphi(t)$ are zero centered,
$\langle \xi(t)\rangle=\langle \varphi(t)\rangle=0$, and the moments of $p=p(0)$ and $q=q(0)$ are
given by Eq.  (\ref{moments}). Taking that into account, squaring and averaging of Eq. (\ref{sol3}) yields
 \begin{eqnarray}
   \langle \dot p(t)^2\rangle=m_s\,T_0 \,[\dot R(t)]^2+
   \frac{m_s\,T_0}{\omega_s^2} \,[\ddot R(t)]^2+\langle \xi^2(t)\rangle+\langle \varphi^2(t)\rangle
 +2\,\langle \xi(t)\,\varphi(t)\rangle.
 \label{aux11}
 \end{eqnarray}
 Here the second moment $\langle \xi^2(t)\rangle$ of the Langevin force is given by Eq. (\ref{xi_moment}),
 so what remains to evaluate in the above equation is the last two terms.
 
 Consider the second moment of $\varphi(t)$,
\begin{eqnarray}
\langle \varphi^2(t)\rangle=\int_0^t d\tau_1
\dot R(t-\tau_1)\int_0^t d\tau_2
\dot R(t-\tau_2)\,\langle\xi_1(\tau)\,
\xi(\tau_2)\rangle.
\label{phi2}
\end{eqnarray}
Using the fluctuation-dissipation relation (\ref{FDR}), we can write this expression as a sum of three terms
\begin{eqnarray}
\langle \varphi^2(t)\rangle&=&
B_1(t)+B_2(t)+B_3(t),
\label{phi3}\\
B_1(t)&=&m_s\,T\,\int_0^t d\tau_1
\dot R(t-\tau_1)\int_0^t d\tau_2
\dot R(t-\tau_2)\,K( |\tau_1-\tau_2|),
\label{B1}\\
B_2(t)&=&-\frac{m_s\,T}{\omega_s^2}\,
[(\dot R*K)(t)]^2,
\label{B2}\\
B_3(t)&=&-\frac{m\,T}{\omega_s^4}\,[(\dot R*\dot K) (t)]^2.
\label{B3}
\end{eqnarray}
Here, as before,  the symbol $*$ stands for a  convolution. Consider first the function $B_1(t)$,
\begin{eqnarray}
B_1(t)=
m_s\,T\,\int_0^t d\tau_1
\dot R(\tau_1)\int_0^t d\tau_2
\dot R(\tau_2)\,K( |\tau_1-\tau_2|)=2m_s\,T\,\int_0^t d\tau_1
\dot R(\tau_1)\int_0^{\tau_1} d\tau_2
\dot R(\tau_2)\,K(\tau_1-\tau_2).
\label{B1_2}
\end{eqnarray}
The inner integral in the last expression is the convolution $(\dot R*K)(\tau_1)$. From  Eq. (\ref{resolvent2}) 
one finds
\begin{eqnarray}
(\dot R*K)(t)=-\ddot R(t)-K(t),
\label{convolution2}
\end{eqnarray}
then
\begin{eqnarray}
B_1(t)=
-2m_s\,T\,\int_0^t 
\dot R(\tau)\,\ddot R(\tau)\,d\tau
-2m_s\,T\int_0^t \dot R(\tau)\, K(\tau)\,d\tau,
\end{eqnarray}
or, taking into account that $\dot R(0)=0$,
\begin{eqnarray}
B_1(t)=
-m_s\,T\,[\dot R(t)]^2-2 m_s T
\int_0^t 
\dot R(\tau)\, K(\tau)\, d\tau.
\end{eqnarray}
The second term $B_2(t)$, with the help of
Eq. (\ref{convolution2}), can be worked out to the form 
\begin{eqnarray}
B_2(t)=-\frac{m_s\,T}{\omega_s^2}\,
[\ddot R(t)+K(t)]^2.
\label{B2_2}
\end{eqnarray}
Expression (\ref{B3}) for the term $B_3(t)$ involves the convolution $(\dot R*\dot K)(t)$. By differentiating Eq. (\ref{resolvent2}) 
twice  and using integration by parts one can get
\begin{eqnarray}
(\dot R*\dot K)(t)=-\dddot R(t)-\omega_s^2\,\dot R(t)-\dot K(t).
\label{convolution3}
\end{eqnarray}
Alternatively, this relation can be derived evaluating the convolution $(\dot R*\dot K)$ in the
Laplace domain with the help of Eq. (\ref{ddotR}).  With Eq. (\ref{convolution3}), 
$B_3(t)$ takes the form 
\begin{eqnarray}
B_3(t)=-\frac{m\,T}{\omega_s^4}\,[
\dddot R(t)+\omega_s^2\,\dot R(t)+\dot K(t)
]^2.
\label{B3_2}
\end{eqnarray}
Substituting the above expressions for $B_1$, $B_2$ and $B_3$ into Eq. (\ref{phi3}) yields
\begin{eqnarray}
\langle\varphi^2(t)\rangle=&-&m_s T\, [\dot R(t)]^2-
\frac{m_s T}{\omega_s^2}\,\left[\ddot R(t)+K(t)\right]^2-
\frac{ m T}{\omega_s^4}\,\left[
\dddot R(t)+\omega_s^2\,\dot R(t)+\dot K(t)\right]^2\nonumber\\
&-&2 m_sT\int_0^t \dot R(\tau)\, K(\tau)\, d\tau.
\label{phisquared}
\end{eqnarray}

Let us  now  evaluate  the last term in Eq. (\ref{aux11}),
\begin{eqnarray}
 2\,\langle \xi(t)\,\varphi(t)\rangle=
 2\int_0^t d\tau\,\dot R(t-\tau)\,\langle \xi(\tau)\,\xi(t)\rangle.
 \label{xiphi}
 \end{eqnarray}
 Using the fluctuation-dissipation relation (\ref{FDR}), one gets
\begin{eqnarray}
 2\,\langle \xi(t)\,\varphi(t)\rangle=
 2m_s\,T\int_0^t \dot R(t-\tau)\,K(t-\tau)\, d\tau-
 \frac{2 m_s\,T}{\omega_s^2}\, (\dot R *K)(t)\, K(t)-
 \frac{2m\,T}{\omega_s^4}\, (\dot R *\dot K)(t)\, \dot K(t).
 \label{xiphi2}
 \end{eqnarray}
Changing variables in the first term and using Eqs. (\ref{convolution2}) and (\ref{convolution3}) for
convolutions $(\dot R*K)$ and $(\dot R*\dot K)$ yields
\begin{eqnarray}
 2\,\langle \xi(t)\,\varphi(t)\rangle&=&
 2m_s\,T\int_0^t \dot R(\tau)\,K(\tau)\, d\tau\nonumber\\
 &+&
 \frac{2 m_s\,T}{\omega_s^2}\, 
 \left[\ddot R(t)+K(t)\right]\,
 K(t)+
 \frac{2m\,T}{\omega_s^4}\,
 \left[\dddot R(t)+\omega_s^2\,\dot R(t)+\dot K(t)\right]\,
 \dot K(t).
 \label{xiphi3}
 \end{eqnarray}

Substituting Eqs. (\ref{xi_moment}), (\ref{phisquared}), and (\ref{xiphi3}) into Eq. (\ref{aux11}) one finds
\begin{eqnarray}
\langle\dot p(t)^2\rangle=m_s\,\omega_s^2\,T+
m_s\,(T_0-T)\,\left[\dot R(t)^2+\frac{1}{\omega_s^2}\,\ddot R(t)^2\right]-\frac{m\,T}{\omega_s^4}\,\left[ \omega_s^2\,\dot R(t)+\dddot R(t)\right]^2.
\end{eqnarray}
Then, according to (\ref{V0}), the average potential energy of the system is
\begin{eqnarray}
V(t)=\frac{T}{2}+\frac{T_0-T}{2}\,
\left\{\omega_s^{-2}\,\dot R(t)^2+
\omega_s^{-4}\,\ddot R(t)^2\right\}
-\frac{\alpha\,T}{2}\,\left\{
\omega_s^{-1}\,\dot R(t)+\omega_s^{-3}\,\dddot R(t)\right\}^2.
\label{V}
\end{eqnarray}
Remarkably, this result can be obtained from expression (\ref{E}) for the average kinetic energy $E(t)$
by making in the latter  the replacement $R(t)\to \omega_s^{-1}\dot R(t)$.

\section{Internal energy}
Combining findings of the previous two sections, i.e. adding up Eq. (\ref{E}) for the average kinetic energy $E(t)$
and Eq. (\ref{V}) for potential energy $V(t)$, for the total internal energy of the system $U(t)=E(t)+V(t)$ we obtain the following result
\begin{eqnarray}
U(t)=T+(T_0-T)\,\psi_1(t)-
\alpha\,T\,\psi_2(t),
\label{U}
\end{eqnarray}
where dimensionless functions  $\psi_1(t)$ and $\psi_2(t)$ are
\begin{eqnarray}
\psi_1(t)&=&\frac{1}{2}\,\left[R^2(t)+2\,\omega_s^{-2}\,\dot R(t)^2+\omega_s^{-4}\,\ddot R(t)^2\right],\nonumber\\
\psi_2(t)&=&
\frac{1}{2}\,\left[R(t)+\omega_s^{-2}\,\ddot R(t)\right]^2+
\frac{1}{2}\left[\omega_s^{-1}\,\dot R(t)+
\omega_s^{-3}\,\dddot R(t)\right]^2.
\label{psi}
\end{eqnarray}
Since $R(0)=1$, $\ddot R(0)=-\omega_s^2$, $\dot R(0)=\dddot R(0)=0$, the initial values of the functions are
\begin{eqnarray}
\psi_1(0)=1, \qquad \psi_2(0)=0,
\label{psi_initial}
\end{eqnarray}
and therefore the initial value of the internal 
energy is 
\begin{eqnarray}
U(0)=T_0.
\label{U_initial}
\end{eqnarray}
This is consistent with our model's assumption that at $t<0$ the system is equilibrated with an external bath at temperature $T_0$. 
The behavior of $U(t)$ at long times is governed by asymptotic properties of the resolvent and its derivatives.
Namely, if the resolvent and its first three derivatives vanish at long times, then so do $\psi_1(t)$ and $\psi_2(t)$,
\begin{eqnarray}
\psi_1(t),\, \psi_2(t)\to 0, \qquad t\to\infty.
\end{eqnarray}
In that case it follows from  Eq. (\ref{U}) that the system is ergodic, i.e. thermalizes with the bath at temperature $T$,
\begin{eqnarray}
U(t)\to T,\qquad \mbox{as}\,\,\,t\to\infty.
\end{eqnarray}
This situation is what we called in the Introduction {\it Scenario 1}. 
On the other hand, if the resolvent and its derivatives do not vanish at long time, then 
it follows from the above relations that the system is not ergodic, i. e. $U(t)$ does not converge to $T$.
Such  situation was referred  to in the Introduction as {\it Scenario 2}.

At a given time the direction and magnitude of heat transfer between the system and bath
is characterized by the change of the internal energy of the system
\begin{eqnarray}
\Delta U(t)=U(t)-U(0)=U(t)-T_0.
\label{DU0}
\end{eqnarray}
From Eqs. (\ref{U}). (\ref{U_initial}), and (\ref{DU0}) we find 
\begin{eqnarray}
\Delta U(t)=(T-T_0)\,[1-\psi_1(t)]-\alpha\,T\,\psi_2(t).
\label{DU}
\end{eqnarray}
Because of the term $-\alpha\,T\psi_2(t)$,
this expression is manifestly in disagreement with the zeroth law of thermodynamics: the energy exchange
between the system and bath is not identically zero when $T=T_0$. If the system is ergodic,  $\psi_2(t)$ vanishes at long times,
and the heat transfer between the system and bath at the same temperature is a transient process,
not observable on the macroscopic time scale.
On the other hand, if the system is non-ergodic, $\psi_2(t)$ does not vanish and the heat exchange between
the system and bath does not respect the zeroth law on any time scale.

The limitation of the zeroth law for the present model is remarkable but hardly a surprise. For a non-ergodic
system, which does not reach thermal equilibrium,  the zeroth law does not apply  anyway. For an ergodic system
the violation occurs on a microscopic time scale, i.e. beyond the application range of macroscopic thermodynamics.
As  we mentioned before, from the point of view of statistical mechanics, 
the transient heat exchange between the system and bath at $T=T_0$ is to 
be expected because the 
the initial distribution $\rho=\rho_s\rho_b$ for the given setup does not take into account 
the system-bath interaction and therefore is not an equilibrium distribution for the overall system even if $T=T_0$.
From this perspective, one may say that the  term $-\alpha\,T\,\psi_2(t)$ in Eq. (\ref{DU}) describes effects
of the strong coupling between the system and bath. 
Note that we did not absorb the factor $\alpha$ in the definition of function $\psi_2(t)$ in order to make it more visible that
in our model effects of the strong coupling are  linear in $\alpha$.  
In the Brownian limit $\alpha\gg 1$, i.e. when the system is much heavier than atoms of the bath, such effects are  small.


It is clear that expression (\ref{DU})  for $\Delta U(t)$ is in general inconsistent 
with the Clausius statement that heat goes from hot to cold.
Indeed, if $T$ and $T_0$ are sufficiently close, then the first term in Eq. (\ref{DU}) is small, 
and the sign of $\Delta U(t)$, and therefore the direction of heat transfer, 
is determined by the strong coupling  term $-\alpha\,T\, \psi_2(t)$, which does not depend on the temperature difference.
Here again the disagreement with macroscopic thermodynamics emerges as a result of the strong coupling of the system and bath.

In order to find precise conditions and time intervals of validity of the 
Clausius statement 
we need to evaluate  $\Delta U(t)$
as an explicit function of time. That requires 
to evaluate 
the resolvent $R(t)$ 
and functions $\psi_1(t)$ and $\psi_2(t)$
in explicit forms.
Recall  that the resolvent $R(t)$ is defined in the Laplace domain by relation (\ref{resolvent1}), 
$\tilde R(s)=1/[s+\tilde K(s)]$.
With the transform of the memory kernel $\tilde K(s)$ given by Eq. (\ref{KK}) one gets
\begin{eqnarray}
\tilde R(s)=\frac{(2-\beta)\,s+\beta\sqrt{s^2+\omega_0^2}}{(2-\beta)\,s^2+\beta\, s\,\sqrt{s^2+\omega_0^2}+2\,\omega_s^2},
\label{R}
\end{eqnarray}
where, recall, $\omega_0^2=4\,k/m$ and $\omega_s^2=k_s/m_s=\alpha\,\beta\,\omega_0^4/4$. 
The inversion of a transform of this form was discussed, for instance,  in Appendix D of Ref.~\cite{MM}. 
In this paper, 
instead of inverting transform (\ref{R})
for  arbitrary $\alpha$ and $\beta$, 
we prefer to focus on 
two specific cases $\beta=1$ and $\beta=2$ (with $\alpha $ being arbitrary),
for which the results  
are more compact and reflect all relevant physics, covering both ergodic and non-ergodic systems.

\section{Resolvent for $\beta=1$}

In the case $\beta=k_s/k=1$ all springs of the overall system are the same and the only parameter of
the model is the mass ratio $\alpha=m/m_s$. The transform of the resolvent  (\ref{R}) takes the form
\begin{eqnarray}
\tilde R(s)&=&\frac{s+\sqrt{s^2+\omega_0^2}}{s^2+ s\,\sqrt{s^2+\omega_0^2}+\alpha\,\omega_0^2/2}.
\label{R1}
\end{eqnarray}
Factorizing the denominator
\begin{eqnarray}
  s^2+ s\,\sqrt{s^2+\omega_0^2}+\frac{\alpha\,\omega_0^2}{2}=\frac{1}{2}\,
  \left(s+\sqrt{s^2+\omega_0^2}\right)\left(\alpha\,\sqrt{s^2+\omega_0^2}+(2-\alpha)\,s\right),
\end{eqnarray}
the above expression is further simplified to 
\begin{eqnarray}
\tilde R(s)&=&\frac{1}{(\alpha/2)\,\sqrt{s^2+\omega_0^2}+(1-\alpha/2)\,s}.
\label{R11}
\end{eqnarray}
If we replace in this expression $\alpha/2\to \alpha$,
it would coincide with the familiar result for the normalized equilibrium correlation function
$C(t)=\langle p(t)\,p(0)\rangle/\langle p^2\rangle$ (here the average is taken with the equilibrium
canonical distribution for the overall system) for an isotope atom in a otherwise uniform harmonic
infinite chain~\cite{Rubin},
\begin{eqnarray}
\tilde C(s)=\frac{1}{\alpha\,\sqrt{s^2+\omega_0^2}+(1-\alpha)\,s}.
\label{C}
\end{eqnarray}
The inverse transform of expression (\ref{C}) 
is well-known
~\cite{Rubin} (see  also Appendix B of paper~\cite{Plyukhin} for technical details), so we can use it 
for the inversion of (\ref{R11}) just replacing 
$\alpha\to\alpha/2$. 
Closed-form expressions for $R(t)$ are available only for $\alpha=2$ and $\alpha=1$,
\begin{eqnarray}
R(t)=\begin{cases}J_0(\omega_0 t),  & \text{if}\quad \alpha=2,
               \\
               \frac{2}{\omega_0 t}\,J_1(\omega_0 t), & 
               \text{if}\quad \alpha=1. 
            \end{cases}
\label{R1_explicit}
\end{eqnarray}
For other values $\alpha<2$ the resolvent can be presented  in the integral form
\begin{eqnarray}
  R(t)=\frac{\alpha}{\pi}\, \int_0^{\omega_0} \frac{\sqrt{\omega_0^2-\omega^2}\, \cos(\omega t)}
  {(1-\alpha)\,\omega^2+\alpha^2\,\omega_0^2/4}\,d\omega.
\label{R1_implicit}
\end{eqnarray}
The resolvent
$R(t)$ of the forms (\ref{R1_explicit}) and (\ref{R1_implicit}) vanishes at long time. Therefore
the system demonstrates an ergodic behavior when $\alpha\le 2$, i.e. when the system's mass is
larger than the half-mass of the bath's atoms,  $m_s\ge m/2$.

For $\alpha>2$ a remarkable phenomenon of a localized vibration occurs~\cite{Montroll,Rubin}.
The resolvent takes the form 
\begin{eqnarray}
R(t)=A(\alpha)\, \cos (\omega_*t)+R_0(t), \qquad \alpha>2.
\label{R_loc}
\end{eqnarray}
Here the function $R_0(t)$ is given 
by the right-hand side of 
Eq. (\ref{R1_implicit}) and vanishes at long times, while 
the frequency and amplitude of the localized vibrational mode
are
\begin{eqnarray}
\omega_*(\alpha)=\frac{\alpha}{2\sqrt{\alpha-1}}\,\omega_0,\qquad
A(\alpha)=\frac{\alpha-2}{\alpha-1}.
\label{omega_loc}
\end{eqnarray}
Thus, for $\alpha>2$ the resolvent does not vanish at long time but oscillates with frequency $\omega_*$.
The system is non-ergodic, it does not reach equilibrium with the bath.
According to Eqs. (\ref{U}) and (\ref{psi}), the internal energy of the system $U(t)$ oscillates with time,
but  its time-average value $\overline{U}$ takes a stationary value. 
We shall see in the next section that  if $T_0<T$ then the time-averaged energy change
$\overline{\Delta U}=\overline U-U(0)$ may be positive, i.e. the colder system releases heat into a hotter bath,
in contradiction with the Clausius statement.

Mathematically,  a condition of the emergence of a localized vibrational mode with frequency $\omega_*$ in
a harmonic lattice is that the function $\tilde R(s)$ has simple poles $\pm i\omega_*$ located on the imaginary
axis, provided the frequency $\omega_*$ is outside the frequency spectrum of the lattice~\cite{Montroll,Rubin}.
The latter condition implies
 $\omega_*>\omega_0$, because $\omega_0$ has the meaning of the highest frequency of the infinite 
 lattice representing the bath.
Analyzing expression (\ref{R1}) for $\tilde R(s)$ one finds that it has indeed simple poles $\pm i\omega_*$
with frequency $\omega_*$ given by Eq. (\ref{omega_loc}). There is, however, a subtlety at this point.
With $\omega_*$ given by Eq. (\ref{omega_loc}), the condition $\omega_*>\omega_0$
is satisfied  for $\alpha>1$.
From this one may erroneously conclude that the condition of the localized mode is $\alpha>1$, rather than $\alpha>2$.
The puzzle is resolved by noting that the function $\tilde R(s)$ has two branches, and only 
one of them is physically meaningful,
i.e. consistent with the correct initial condition $R(0)=1$. One can show that $s=\pm i\omega_*$
are the pole for the physical branch of $\tilde R(s)$ only
for $\alpha>2$.
For $1<\alpha\le 2$
the function $\tilde R(s)$
still has the poles at $s=\pm i\omega_*$ with $\omega_*>\omega_0$, but they
corresponds to the unphysical branch and thus should be discarded, see Appendix B of Ref.~\cite{Plyukhin} for details.

Now equipped with explicit expressions for the resolvent (although so far only for $\beta=1$),
we can exploit expressions (\ref{U}) and (\ref{psi}) for the internal 
energy to explore the energy exchange between the system and bath.  
We shall consider ergodic and non-ergodic systems separately.

\section{Heat transfer for ergodic system ($\alpha=2$, $\beta=1$)}
As an example of an ergodic system, consider the case $\beta=1$ and $\alpha=2$,
when the resolvent has a simple analytical form  $R(t)=J_0(\omega_0 t)$, see Eq. (\ref{R1_explicit}),
and the system internal frequency is  $\omega_s^2=\alpha\,\beta\,\omega_0^2/4= \omega_0^2/2$.
Substituting of that expressions  into Eqs. (\ref{psi}) we get $\psi_1$ and $\psi_2$ in the following explicit form
\begin{eqnarray}
\psi_1(\tau)&=&\frac{5}{2}\,J_0(\tau)^2+2\,\left(
1+\frac{1}{\tau^2}\right)\, J_1(\tau)^2
-\frac{4}{\tau}\,J_0(\tau)\,J_1(\tau),
\\
\psi_2(\tau)&=&\left(\frac{1}{2}+\frac{4}{\tau^2}\right)\,J_0(\tau)^2+
\left(1-\frac{6}{\tau^2}+\frac{16}{\tau^4}\right)\,J_1(\tau)^2+
\left(\frac{2}{\tau}-\frac{16}{\tau^3}\right)\,
J_0(\tau)\,J_1(\tau).
\end{eqnarray}
Here $\tau=\omega_0 t$, and we have chosen 
to use  the Bessel functions of zeroth and first orders only. These expressions are defined 
for $\tau>0$, and at $\tau=0$ they should be defined by continuity,
\begin{eqnarray}
\psi_1(0)=\lim_{\tau\to 0}\psi_1(\tau)=1, \qquad 
\psi_2(0)=\lim_{\tau\to 0}\psi_2(\tau)=0.
\end{eqnarray}
According to Eq.  (\ref{psi_initial}), those are the correct initial values.

\begin{figure}[t]
\includegraphics[height=9cm]{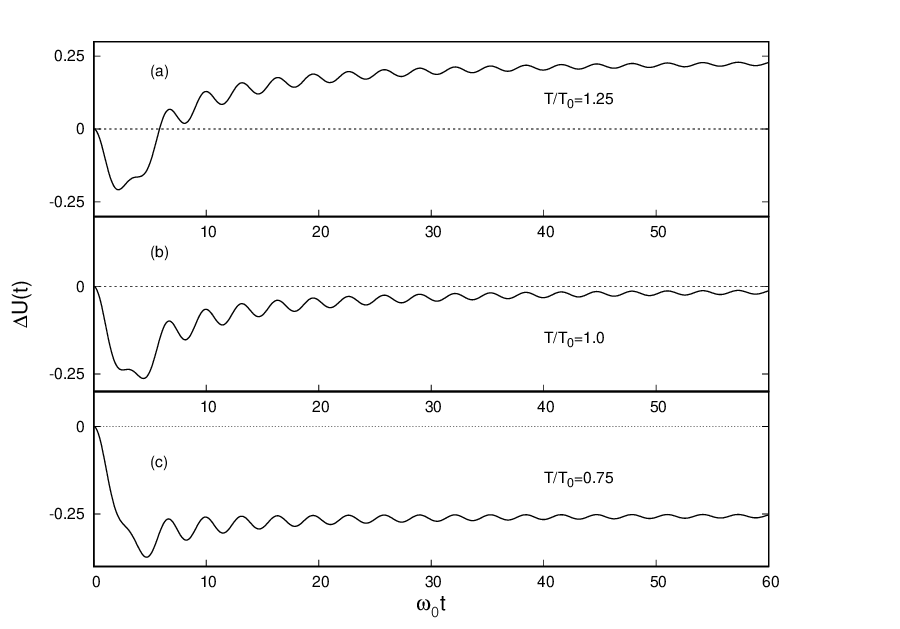}
\caption{ 
The relative change of the internal energy of the system $\Delta U(t)=U(t)-U(0)$, in units $U(0)=T_0$, 
as a function of scaled time for $\alpha=2$ and $\beta=1$, 
for different  values of the temperature ratio $T/T_0$ ($T_0$ is the initial temperature of the system, 
$T$ is temperature of the bath).   
}
\end{figure}

Substituting the above expressions for $\psi_1(t)$ and
$\psi_2(t)$ into Eq. (\ref{DU}),
\begin{eqnarray}
\Delta U(t)=U(t)-U(0)=(T-T_0)\,[1-\psi_1(t)]-2\,T\,\psi_2(t),
\label{DU2}
\end{eqnarray}
gives $\Delta U(t)$ as an explicit function of time. Fig. 2
shows the result for three values of the temperature ratio $T/T_0$.
In all three cases,  the energy change converges at long times to the value $\Delta U_\infty=T-T_0$,
which is consistent with the Clausius statement: the colder system absorbs heat from 
the hotter bath ($\Delta U_\infty>0$), the hotter system releases heat into the colder bath ($\Delta U_\infty<0$),
and the net heat exchange is null when the temperatures of the system and bath are the same.

However, 
one observes from Fig. 2 that while the  Clausius statement holds on
the asymptotically long time scale,
at short times it does not. The function $\Delta U(t)$ is not monotonic; on short time
intervals it increases/decreases regardless of whether the system hotter or colder than the bath.
Such behavior is what was referred to  as {\it Scenario 1} in Introduction. In particular,
one observes from Fig. 2 that regardless of the sign of the temperature difference
$T-T_0$ the system initially loses energy: the function $\Delta U(t)$ first decreases, reaches an absolute  
minimum, and then on a much longer time scale approaches non-monotonically the equilibrium value from below.
Such an {\it initial transient cooling} (the term is suggested by the referee) may be interpreted as  a result of 
the initial energy transfer from the system to the boundary atom of the bath. The latter is initially fixed, see Fig. 1,
and immediately after being released  at $t=0$  it is always ``colder" than the
system, even if the bath's temperature is higher than that of the system.
The  net energy balance results from the interplay of two processes. The first process is the system
releasing heat to to the colder boundary atom, the second process  is the system absorbing heat from the hotter bath.
The initial transient cooling may be viewed as the result that the first process dominates on a shorter time scale.
The second process dominates on the longer time scale, and one expects the transient cooling to be more conspicuous
when the second process is weaker, i.e. when the temperature of the bath is lower.
This trend is visible in Fig. 2: for a fixed initial temperature of the system $T_0$, the extent of the initial transient
cooling increases when the temperature of the bath $T$ decreases.  Instead, at higher $T$ 
one expects the transient cooling to be unimportant.
Indeed, plotting $\Delta U(t)$ according to  Eq. (\ref{DU2}) 
one finds that the initial transient cooling is practically invisible for $T/T_0\ge 10$.

\section{Heat transfer for non-ergodic system ($\alpha>2$, $\beta=1$)}
As discussed in Sec. X, 
for $\beta=1$ and $\alpha>2$ the system shows non-ergodic behavior due to formation of
the localized vibrational mode.
The resolvent is given by Eq. (\ref{R_loc}), $R(t)=A\,\cos\omega_* t+R_0(t)$.
At long times the function $R_0(t)$ vanishes, and the resolvent oscillates
\begin{eqnarray}
R(t)\approx A\,\cos(\omega_* t)
\end{eqnarray}
with the amplitude and frequency given by Eqs. (\ref{omega_loc}).
The functions $\psi_1(t)$ and $\psi_2(t)$, given by Eqs. (\ref{psi}), take the  forms
\begin{eqnarray}
\psi_1(\tau)&=&\frac{A^2}{2}\,\left[1+\left(\frac{\omega_*}{\omega_s}\right)^4\right]\,\cos (\omega_* t)^2+A^2\left(
\frac{\omega_*}{\omega_s}\right)^2\sin(\omega_* t)^2,
\\
\psi_2(\tau)&=&
\frac{A^2}{2}\,\left[1-\left(\frac{\omega_*}{\omega_s}\right)^2\right]^2\,\left[\cos (\omega_* t)^2+
\left(\frac{\omega_*}{\omega_s}\right)^2\,\sin (\omega_* t)^2
\right].
\end{eqnarray}
The time averages of these expressions, which we denote with the overbar,  are
\begin{eqnarray}
\overline{\psi_1}=\frac{A^2}{4}\,\left[1+\left(\frac{\omega_*}{\omega_s}\right)^2\right]^2,
\qquad
\overline{\psi_2}=
\frac{A^2}{4}\,\left[1-\left(\frac{\omega_*}{\omega_s}\right)^2\right]\,\left[1-
\left(\frac{\omega_*}{\omega_s}\right)^4
\right].
\label{psi_av0}
\end{eqnarray}
Taking into account that $A=(\alpha-2)/(\alpha-1)$ and
$(\omega_*/\omega_s)^2=\alpha/(\alpha-1)$, see Eqs. (\ref{omega_loc}) and (\ref{freq_s}), we can express 
the above expressions as functions of the mass ratio $\alpha$ as follows
\begin{eqnarray}
\overline{\psi_1}(\alpha)=
\frac{(\alpha-2)^2 \,(2\alpha-1)^2}{4\,(\alpha-1)^4},
\qquad
\overline{\psi_2} (\alpha)=
\frac{(\alpha-2)^2\, (2\alpha-1)}{4\,(\alpha-1)^5}.
\label{psi_av}
\end{eqnarray}
The plots of the functions $\overline{\psi_1}(\alpha)$ and 
$\alpha\,\overline{\psi_1}(\alpha)$, as well as another  relevant function $\gamma(\alpha)$ defined below,
are shown in Fig. 3.

According to Eq. (\ref{U}),
the time-average change of the system's internal energy is
\begin{eqnarray}
\overline{\Delta U}=\overline{U}-U(0)=(T-T_0)\,[1-\overline{\psi_1}]-T\,\alpha\,\overline{\psi_2}.
\label{DU3}
\end{eqnarray}
The direction of the time-average heat transfer is determined by the sign of this expression.  
The heat transfer is anomalously 
directed (non-Clausius) if 
$\overline{\Delta U}<0$
when the system is initially colder than bath ($T-T_0>0$), or if  $\overline{\Delta U}>0$ when
the system is initially hotter than bath  ($T-T_0<0$). It is easy to see that 
the latter case actually does not occur for the present model.
Indeed, the inequality $\overline{\Delta U}>0$ can be written as 
\begin{eqnarray}
(T-T_0)(1-\overline{\psi_1})>T\,\alpha\,\overline{\psi_2},
\end{eqnarray}
If $T-T_0<0$, it has no solutions because for any  
$\alpha>2$
the left-hand side is negative  and the right-hand side is positive (note that $\overline{\psi_1}(\alpha)<1$
and  $\overline{\psi_2}(\alpha)>0$, see Fig. 3). Thus, if the system is initially hotter than the bath, the heat
transfer is in agreement with the Clausius statement, i.e. is directed from the hotter system to colder bath.

\begin{figure}[t]
\includegraphics[height=7cm]{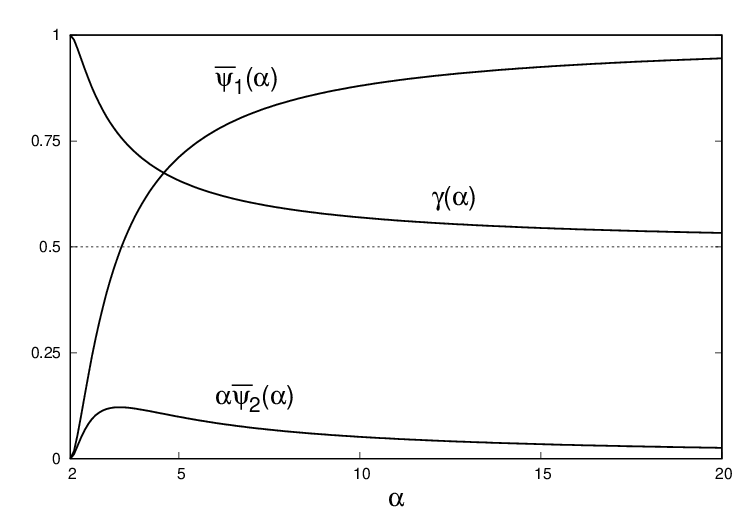}
\caption{ 
Dimensionless functions $\overline{\psi_1}(\alpha)$,
$\alpha\,\overline{\psi_2}(\alpha)$, and  $\gamma(\alpha)$, defined by Eqs. (\ref{psi_av}) and (\ref{gamma}), which determine
the time-average heat exchange  of a non-ergodic system
for the case $(\alpha>2,\beta=1)$, as discussed in Sec. XII. 
}
\end{figure}

The situation is more interesting when the system is initially colder than the bath, $T-T_0>0$. For that case the Clausius
statement suggests that the system  absorbs heat from the bath, so that $\overline{\Delta U}>0$.
However, 
solving the inequality 
\begin{eqnarray}
\overline{\Delta U}=(T-T_0)\,[1-\overline{\psi_1}]-T\,\alpha\,\overline{\psi_2}>0
\label{DU4}
\end{eqnarray}
one finds that the Clausius transfer only occurs
if the system's initial temperature $T_0$ is not too high, namely
\begin{eqnarray}
T_0<T_c=\gamma(\alpha)\, T,
\label{Tc}
\end{eqnarray}
where the function $\gamma(\alpha)$ is
\begin{eqnarray}
{\gamma}(\alpha)=
1-\frac{\alpha\,\overline{\psi_2}(\alpha)}{1-\overline{\psi_1}(\alpha)}=
\frac{2\,\alpha^3-4\,\alpha^2+\alpha}{4\,\alpha^3-13\,\alpha^2+13\,\alpha-4}.
\label{gamma}
\end{eqnarray}
For $\alpha>2$ the function $\gamma(\alpha)$ monotonically decreases from 
from $\gamma(2)=1$ to $\gamma=0.5$ at asymptotically large $\alpha$, see Fig. 3. Therefore, for any $\alpha>2$
the critical temperature $T_c$  is lower than temperature of the bath, 
but bounded from below by the half-temperature of the bath, 
\begin{eqnarray}
T/2<T_c<T.
\end{eqnarray}
If the initial temperature $T_0$ of the system is in the 
interval
\begin{eqnarray}
T_c<T_0<T
\end{eqnarray}
then one finds 
\begin{eqnarray}
\overline{\Delta U}=(T-T_0)\,[1-\overline{\psi_1}]-T\,\alpha\,\overline{\psi_2}<0,
\end{eqnarray}
which corresponds to the non-Clausius heat transfer from
the initially  colder system to hotter bath.
The anomalous heat transfer from the system to bath
also occurs when their temperatures are the same, 
in which case  $\overline{\Delta U}=-T\,\alpha\,\overline{\psi_2}<0$.

These results  may be interpreted as follows. Due to the formation of a localized vibrational mode,
the system exchanges heat not with the entirety of the  bath, but only with a finite fragment
of the bath adjacent to the system.  
Suppose one wishes  to introduce an effective local temperature of that fragment.
Clearly, it must be lower than the bath's bulk  temperature $T$ because
the fragment includes the initially frozen boundary atom $i=0$. It is tempting to  identify the fragment's
effective temperature  with the critical temperature $T_c$ defined by (\ref{Tc}). Then our results indicate that
the Clausius statement 
breaks down when applied to the whole bath, yet
is still valid when applied to the heat exchange between the system and the boundary fragment of the bath provided the
latter has an effective temperature $T_c$: Heat is transferred from the system to the fragment if 
the system's temperature is higher than temperature of the fragment $T_0>T_c$ (though perhaps lower than the bulk
temperature of the bath $T_0<T$), and in the opposite direction otherwise.

\section{Heat transfer for non-ergodic system ($\beta=2$)}
In the previous two sections we discussed the model for $\beta=1$, which shows both ergodic (for $\alpha\le 2$)
and non-ergodic (for $\alpha>2$) behavior. The condition of non-ergodicity $\alpha>2$ implies that the isotope
representing the system is at least twice lighter than atoms of the bath. In contrast, for $\beta=2$ a localized
mode  emerges, and the system is non-ergodic,  for {\it any} value of the mass ratio $\alpha$, including
the Brownian limit $\alpha\ll 1$.
This peculiar feature is the incentive to consider the case $\beta=2$ in this section as our second showcase example.
We shall see, however, that the results for $\beta=2$ are qualitatively similar
to those for the case $(\alpha>2,\, \beta=1)$ discussed in the previous section.

We have seen in Sec. V that for $\beta=2$ the memory kernel in the Langevin equation
takes a simple form, $K(t)=(\alpha\,\omega_0^2/2)\, J_0(\omega_0 t)$.  However, this simplicity does not offer any
particular advantage for the evaluation of the resolvent $R(t)$.
The general expression for the resolvent's transform (\ref{R}) for $\beta=2$ reads
\begin{eqnarray}
\tilde R(s)=\frac{\sqrt{s^2+\omega_0^2}}
{s\,\sqrt{s^2+\omega_0^2}+\alpha\,\omega_0^2/2}
\label{r1}
\end{eqnarray}
and cannot be inverted in terms of standard functions.
One has to be aware of a subtlety related to this expression: If one tries to evaluate the long time limit of the resolvent
using the final value theorem one  gets zero,
\begin{eqnarray}
\lim_{t\to\infty} R(t)=\lim_{s\to 0} s\,\tilde R(s)=0, 
\label{limit}
\end{eqnarray}
which suggests ergodicity. Actually, result (\ref{limit}) is incorrect because, as we shall see,  the long-time
limit of the function $R(t)$ with transform (\ref{r1}) for any positive $\alpha$ does not exist, 
and the final value theorem cannot be applied.

The inversion of transform (\ref{r1}) is discussed in detail in Appendix B. It is
similar to that for the case $\beta=1$,  but also involves some peculiar details.  As was mentioned above,
the inversion is not a merely  mathematical exercise because the function $\tilde R(s)$ has two branches,
and one has to be careful to chose a physically meaningful branch.  
Function (\ref{r1}) has four simple poles, but only two of them are on the physical branch.
Those two are located on the imaginary axis and have the form
$s=\pm i\,\omega_*$ where
\begin{eqnarray}
 \omega_*=\sqrt{
 \frac{1+\sqrt{1+\alpha^2}}{2}
 }\,\omega_0.
\label{w_loc2}
\end{eqnarray}
For any value of the mass ratio $\alpha$, the frequency $\omega_*$ is higher than $\omega_0$ and thus
lies outside the spectrum of the bath's normal modes.  This is just the  condition of the localized mode formation.
The detailed calculation (see Appendix B) gives for the resolvent the result structurally similar
to the one for the case $(\beta=1,\,\alpha>2)$
\begin{eqnarray}
R(t)=R_0(t)+A\,\cos\omega_* t
\end{eqnarray}
where the function $R_0(t)$ is now defined by the integral
\begin{eqnarray}
R_0(t)=
\frac{\alpha\,\omega_0^2}{\pi} \int_{0}^{\omega_0}
\frac{\cos(\omega\,t)\,\sqrt{\omega_0^2-\omega^2}}{\alpha^2\omega_0^4/4+\omega^2\,(\omega_0^2-\omega^2)}\,d\omega.
\end{eqnarray}
As for the case $(\beta=1,\,\alpha>2)$, at long times $R_0(t)$  vanishes, and the resolvent oscillates
$R(t)\approx A\,\cos\omega_* t$
with the frequency $\omega_*$ given by (\ref{w_loc2}) and the amplitude
\begin{eqnarray}
A=1-\frac{1}{\sqrt{1+\alpha^2}}.
\label{AA3}
\end{eqnarray}
The internal energy change of the system $\Delta U(t)$ also oscillates at long times.
Its time average $\overline{\Delta U}$ is given by the same expression 
(\ref{DU3}) 
as for the case $(\beta=1, \,\alpha>2)$,
\begin{eqnarray}
\overline{\Delta U}=\overline{U}-U(0)=(T-T_0)\,[1-\overline{\psi_1}]-T\,\alpha\,\overline{\psi_2}, 
\label{DU44}
\end{eqnarray}
where the time-averaged functions $\overline{\psi_1}$
and $\overline{\psi_1}$ are still given  by Eqs. (\ref{psi_av0}). For $\beta=2$
the squared internal frequency of the system 
is $\omega_s^2=\alpha\,\omega_0^2/2$, and 
\begin{eqnarray}
\left(\frac{\omega_*}{\omega_s}\right)^2
=\frac{
1+\sqrt{1+\alpha^2}
}{\alpha}.
\end{eqnarray}
Substituting this  and  Eq. (\ref{AA3}) for  $A$ into  Eqs. (\ref{psi_av}) yields for  
$\overline{\psi_1}$
and $\overline{\psi_1}$ as explicit functions
of $\alpha$ the following expressions
\begin{eqnarray}
\overline{\psi_1}(\alpha)=
\frac{
(\sqrt{\alpha^2+1}+\alpha-1)^2
}{4\,(\alpha^2+1)},
\qquad
\overline{\psi_2} (\alpha)=\frac{
\alpha+1-\sqrt{\alpha^2+1}
}{2\,(\alpha^2+1)}.
\label{psi_av3}
\end{eqnarray}
While these expressions are different than those for the case $(\beta=1,\,\alpha>2)$,
the qualitative behavior of functions $\overline{\psi_1}(\alpha)$
and $\overline{\psi_2}(\alpha)$
in two cases is similar, see  Fig. 3, except that the functions
domain in the present case $\beta=2$ is $\alpha>0$ instead of $\alpha>2$.

Repeating the analysis of Section XII, i.e. solving inequalities $\overline{\Delta U}>0$ and $\overline{\Delta U}<0$
for different signs of the temperature difference $T-T_0$, one finds results similar to the case $(\beta=1,\,\alpha>2)$. 
Namely, 
if the system is initially hotter  than the bath $T_0>T$, then the heat transfer is normal,
i.e. the system loses energy, $\overline{\Delta U}<0$. On the other hand, one finds that
the system may lose energy even if it is initially colder than the bath,
\begin{eqnarray}
\overline{\Delta U}<0, \quad \mbox{for}\quad T_0<T,
\label{auxx}
\end{eqnarray}
i.e. the heat transfer may be non-Clausius, 
provided the system temperature is higher than a  critical temperature $T_c$,
\begin{eqnarray}
T_c<T_0<T.
\end{eqnarray}
Solving inequality (\ref{auxx}), one finds for the critical temperature the expression $ 
T_c=\gamma(\alpha)\, T$
with 
\begin{eqnarray}
\gamma(\alpha)=1-\frac{\alpha\,
\overline{\psi_2}
}{1-\overline{\psi_1}}=\frac{1+\sqrt{\alpha^2+1}}{1+\alpha+\alpha^2+\sqrt{1+\alpha^2}-\alpha\,\sqrt{1+\alpha^2}}.
\end{eqnarray}
The function $\gamma(\alpha)$ behaves in a way  qualitatively similar to  that for the case $(\beta=1,\,\alpha>2)$,
i.e. it monotonically decreases from $1$ to $1/2$.
Thus, we find  for the critical temperature the same lower bound $T_c>T/2$ 
as for the case 
$(\beta=1,\,\alpha>2)$.

Similar to the case $(\alpha>2, \beta=1)$, we can interpret
the results  arguing that the boundary region of the bath is characterized by the effective temperature $T_c$.
Since $T_c$ is lower than the bath's bulk temperature $T$, a non-Clausius heat transfer from the colder system to
hotter bath can be interpreted as a Clausius transfer from the system to the bath's boundary region
when the former is hotter than the latter. 

\section{Conclusion}

Thermodynamics is a macroscopic theory, and 
at present 
there is no consensus on  to what extent and under what conditions it can be extended 
to microscopic and  mesoscopic systems.
Quoting  Ref.~\cite{IO}: {\it The conclusions of thermodynamics apply to macroscopic systems only.
  A system with small number of particles will not obey the laws of thermodynamics, especially the second law.} 
Nevertheless, many efforts and progress have been made in recent years in constructing thermodynamics of small
systems strongly coupled to the environment~\cite{Gelin,Seifert,Jar,Talkner}. In this paper
we have followed a somewhat opposite route studying
conditions when  properties of microscopic open systems may be at odds with macroscopic thermodynamics.

We found that the second law in the form of 
the Clausius statement (heat does not spontaneously flow from cold to hot) 
does not generally hold, yet 
it is quite robust. 
For ergodic systems  we found that the Clausius  statement may be violated on microscopically short time intervals,
yet it still holds on a coarse-grained time scale with a sufficiently low (``macroscopic") time resolution.
In particular, if one measures heat transfer for a transition with initial and final states being equilibrium
ones (which means that the transition occurs on a time scale longer than the thermalization time of the system),
the Clausius statement is valid and in agreement with other forms of the second law. The violation of
the Clausius statement on a time scale shorter than the thermalization time involves the system out of
equilibrium and  does not contradict the second law  in the form of the Clausius inequality, since
the latter refers to transitions with  initial and final states
(but not necessarily  intermediate  states) being equilibrium ones.

Perhaps a more interesting result  is that
the Clausius statement may not hold in any sense for a non-ergodic system,
which does not reach thermal equilibrium with the bath due to the formation of  a localized vibrational mode.
Again, this finding by no means compromises macroscopic thermodynamics, which concerns  ergodic systems only.
Still, we believe that the result is 
of interest as a concrete  example 
which shows limitations of the (simple) thermodynamic description of a (complex) dynamical process of
heat transfer involving small systems. Specifically, we found that 
the anomalously directed heat transfer from a cold non-ergodic system with initial temperature $T_0$
to a hotter bath with temperature $T>T_0$ occurs if the system temperature is higher than a certain
critical temperature $T_c$,
\begin{eqnarray}
T_c<T_0<T.
\label{co1}
\end{eqnarray}
This suggests to interpret $T_c$ as an effective temperature of a fragment of the bath adjacent to the
system and involved in a localized vibrational motion. That fragment, in the studied setup, is effectively
cooler than the rest of the bath because the boundary atom is initially fixed. Then our results are
naturally accounted for by the 
assumption  that the Clausius statement is still valid
if we replace the bulk temperature of the bath $T$ by the bath's local temperature $T_c$ at the boundary: 
Although the system is colder than the bath, $T_0<T$, in the presence of a localized vibrational mode the system
effectively interacts only with a bath's small boundary region with the effective temperature $T_c<T$.
Then the system releases heat into the boundary region if $T_0>T_c$, and absorbs heat from the region if $T_0<T_c$.

The value of $T_c$ depends on parameters of the model $\alpha$ and $\beta$, but in all considered cases
it is found to be bounded from below by the half-temperature of the bath, $T_c>T/2$.

A simpler model of Ref.~\cite{Plyukhin}, where the  system acquires the equilibrium distribution instantaneously,
shows a similar behavior, but the result in that case is reversed in the sense that an 
anomalously directed heat transfer occurs from a colder bath to a hotter non-ergodic system
(instead of from a colder system to a hotter bath in the present model).

In view of these findings, it is  natural to ask whether 
the protocol studied here
 can be used to design a perpetual motion machine of the second kind. It is clear that   
one can use the set up
with a non-ergodic system (e.g. when $\beta=2$, or when $\beta=1,\alpha>2)$
to transfer some (small) amount of energy $\Delta U$ from a colder system to a hotter one.
However, in order to arrange such transfer in a systematic way, we need to return the overall system into
the  initial configuration, depicted in Fig. 1, with the boundary atom fixed at the position corresponding
to the mechanical equilibrium of the chain. Physically, a periodic protocol can be arranged by trapping the
boundary atom in an external potential well, which can turned on and off in proper moments.
This, however,  appears to be impossible without some  Maxwell's demon-like apparatus.

As a technical tool, we derived and exploited  the generalized Langevin equation (\ref{GLE})
with a non-stationary noise. The non-stationarity of the noise reflects the non-stationarity
of the heat transfer in the studied setting. 
The fluctuation-dissipation relation we found, see Eq. (\ref{FDR}), differs from the standard one
by the presence of additional terms involving not only the dissipative kernel, but also the kernel's first derivative.
Although the forms of the Langevin equation and fluctuation-dissipation relation used in this
paper are model-sensitive, we believe they may be of interest as a simple example of the
Langevin dynamics extended beyond the standard assumptions.

Although the linearity of the presented model is essential for all calculations,  we believe
that qualitatively our findings are not specific for linear systems only, in particular, because
a non-ergodic behavior, similar to that considered here, is known to occur in nonlinear systems as well~\cite{Dhar}.

\renewcommand{\theequation}{A\arabic{equation}}
  \setcounter{equation}{0}  

  \section*{APPENDIX A: Derivation of Langevin equation for the boundary atom}  
In this appendix we derive the generalized Langevin equations (\ref{le}) for the boundary atom $i=0$.

According to Eq. (\ref{eom2}), the equation of motion of 
the boundary atom is that of an oscillator linearly coupled to the system and also to normal mode oscillators of the bath,
\begin{eqnarray}
\dot p_0=-k_s\,(q_0-q) -k\,q_0 +\sum_{j=1}^N c_j Q_j.
\label{A1}
\end{eqnarray}
The right part of Fig. 1  gives a pertinent illustration.
Normal mode coordinates $Q_j(t)$ satisfy Eq. (\ref{eom3}),
\begin{eqnarray}
\ddot Q_j=-\omega_j^2\,Q_j+c_j\, q_0,
\label{A2}
\end{eqnarray}
which has the general solution
\begin{eqnarray}
Q_j(t)=Q_j^0(t)+\frac{c_j}{\omega_j}\,\int_0^t \sin\omega_j (t-t')\, q_0(t')\,dt'.
\label{A3}
\end{eqnarray}
Here $Q_j^0(t)$ is a solution of the corresponding homogeneous equation 
\begin{eqnarray}
Q_j^0(t)=Q_j(0)\,\cos \omega_j t+\frac{P_j(0)}{\omega_j}\,\sin \omega_j t.
\label{A4}
\end{eqnarray}
Physically, $Q_j^0(t)$ describes evolution of normal modes when the boundary atom is fixed, $q_0=0$.
Integrating the second term in Eq. (\ref{A3}) by parts and taking into account that $q_0(0)=0$, one gets
\begin{eqnarray}
Q_j(t)=Q_j^0(t)+
\frac{c_j}{\omega_j^2}\,\left\{
q_0(t)
-\frac{1}{m}\,\int_0^t \cos\omega_j(t-t')\,p_0(t')\,dt'
\right\}.
\label{A5}
\end{eqnarray}
Substitution of this expression  into Eq. (\ref{A1}) gives the generalized Langevin equation
\begin{eqnarray}
\dot p_0(t)=-\left\{
k_s+k-\sum_{j=1}^N \left(
\frac{c_j}{\omega_j}\right)^2
\right\}\,q_0(t)+k_s\,q(t)
-\int_0^t K_0(t-t')\,p_0(t')\,dt' +\eta(t).
\label{A6}
\end{eqnarray}
with the fluctuating force 
\begin{eqnarray}
\eta(t)=\sum_{j=1}^N c_j Q_j^0(t).
\label{A7}
\end{eqnarray}
and the memory kernel
\begin{eqnarray}
K_0(t)=\frac{1}{m}\sum_{j=1}^N 
\left(
\frac{c_j}{\omega_j}
\right)^2\cos\omega_j t.
\label{A8}
\end{eqnarray}

Eqs. (\ref{A6})-(\ref{A8}) are exact and hold for any $N$.
They take a more compact form for the  infinite bath, $N\to\infty$.
As follows from  Eqs. (\ref{H_b2}) and (\ref{c}),
\begin{eqnarray}
\left(\frac{c_j}{\omega_j}\right)^2=\frac{2k}{N+1}\,\cos^2\frac{\pi j}{2(N+1)},
\label{A9}
\end{eqnarray}
then one observes that
\begin{eqnarray}
\sum_{j=1}^N
\left(
\frac{c_j}{\omega_j}
\right)^2=k\,\frac{N}{N+1}\to k, \qquad \mbox {as}\,\, N\to \infty.
\end{eqnarray}
Therefore, in the limit $N\to\infty$ the Langevin equation (\ref{A6}) takes the form
\begin{eqnarray}
\dot p_0(t)=-k_s\,[q_0(t)-q(t)]
-\int_0^t K_0(t-t')\,p_0(t')\,dt' +\eta(t),
\label{A11}
\end{eqnarray}
which is just  Eq. (\ref{le}) of the main text. 

With Eqs. (\ref{A9}) and (\ref{H_b2}), expression (\ref{A8}) for the kernel $K_0(t)$ gives
\begin{eqnarray}
K_0(t)=
\frac{\omega_0^2}{2(N+1)}\,\sum_{j=1}^N \cos^2\left(
\frac{\pi}{2}\frac{j}{N+1}\right)\,\cos\left(
\omega_0 t\,\sin\left(\frac{\pi}{2} \frac{j}{N+1}\right)\right),
\end{eqnarray}
where $\omega_0=2\sqrt{k/m}$.
In the limit $N\to\infty$ this expression takes the integral form 
\begin{eqnarray}
K_0(t)=\frac{\omega_0^2}{\pi}
\,\int_0^{\pi/2} \cos^2\theta\,\cos(\omega_0 t\,\sin\theta)\,d\theta,
\end{eqnarray}
which gives for the kernel expression (\ref{K_0}) in terms
of Bessel functions,
\begin{eqnarray}
K_0(t)=\frac{\omega_0^2}{4}\,[J_0(\omega_0 t)+J_2(\omega_0 t)].
\label{A14}
\end{eqnarray}

Using expression (\ref{A7}) for the fluctuating force $\eta(t)$ and distribution (\ref{rho_b})
for initial bath variables, one can verify directly that $\eta(t)$ is zero-centered, stationary,
and related to the kernel $K_0(t)$ by the standard fluctuating-dissipating relation (\ref{fdt_0}).

\renewcommand{\theequation}{B\arabic{equation}}
  \setcounter{equation}{0}  

\renewcommand{\theequation}{B\arabic{equation}}
  \setcounter{equation}{0}  

\section*{APPENDIX B: Evaluation of resolvent $R(t)$ for the case $\beta=2$}  

In this Appendix we present
the inversion of the  Laplace transform (\ref{r1}) 
\begin{eqnarray}
\tilde R(s)=\frac{\sqrt{s^2+\omega_0^2}}{s\sqrt{s^2+\omega_0^2}+\alpha\,\omega_0^2/2}
\label{B_r1}
\end{eqnarray}
of the resolvent $R(t)$ for the case $\beta=2$ and arbitrary positive $\alpha$.

Function (\ref{B_r1}) has two branches which we denote
$\tilde R_1(s)$ and $\tilde R_2(s)$ and write as
\begin{eqnarray}
 \tilde R_k(s)=\frac{f_k(s)}{s\,f_k(s)+\alpha\,\omega_0^2/2}, \qquad k=1,2
 \label{B_branches}
\end{eqnarray}
where $f_1(s)$ and $f_2(s)$ are the two branches of the square-root function 
\begin{eqnarray}
 f(s)=\sqrt{s^2+\omega_0^2}=\sqrt{s+i\omega_0}\,\sqrt{s-i\omega_0}.
\end{eqnarray}
It is convenient to define a branch cut as a segment of the 
imaginary axis connecting the branch points $\pm i\omega_0$, 
and to define $s\pm i\omega_0$ in  a polar form,
\begin{eqnarray}
 s-i\omega_0=r_1\,e^{i\theta_1},\qquad
 s+i\omega_0=r_2\,e^{i\theta_2},
\label{polar}
\end{eqnarray}
see Fig. 4.
Then the two branches of $f(s)$ can be defined by the following expressions:
\begin{eqnarray}
 f_k(s)=\sqrt{r_1\,r_2}\,\, e^{i\,\frac{\theta_1+\theta_2}{2}}, \qquad k=1,2
\label{f12}
\end{eqnarray}
where the ranges of arguments $\theta_1$ and $\theta_2$
for the first branch $f_1(s)$ are  the same,
\begin{eqnarray}
 -\frac{3\pi}{2}<\theta_1\le\frac{\pi}{2},\qquad
-\frac{3\pi}{2}<\theta_2\le\frac{\pi}{2},
 \label{branch1}
\end{eqnarray}
while for the second branch $f_2(s)$ the range of $\theta_2$ is shifted by $2\pi$,
\begin{eqnarray}
 -\frac{3\pi}{2}<\theta_1\le\frac{\pi}{2}, \qquad
 \frac{\pi}{2}<\theta_2\le\frac{5\pi}{2}.
 \label{branch2}
\end{eqnarray}
 
One can verify that the functions $f_1(s)$ and $f_2(s)$
 defined in this way are continuous at any $s$
except on the branch cut.  
In what follows we shall  need to refer  to the following mapping rules for the functions $f_1(s)$ and $f_2(s)$:

(a) Let $s=i\,y$ with $y>\omega_0$ be on the positive imaginary axis above the branch cut. 
Then the first branch  $f_1(s)=\sqrt{r_1\,r_2}\,e^{i\,\pi/2}=i\,\sqrt{r_1\,r_2}$ has a positive imaginary part, 
while the second branch 
$f_2(s)=
\sqrt{r_1\,r_2}\, e^{i\,3\pi/2}=-i\,\sqrt{r_1\,r_2}$
has a negative imaginary part.

(b) Let $s=-i\, y$ with $y>\omega_0$ be on the negative imaginary axis below the branch cut. 
Then the first branch 
$f_1(s)=
\sqrt{r_1\,r_2}\, e^{-i\,\pi/2}=-i\,\sqrt{r_1\,r_2}$ has a negative imaginary part, while the second branch
 $f_2(s)=
\sqrt{r_1\,r_2}\, e^{i\,\pi/2}=i\,\sqrt{r_1\,r_2}$
has a positive imaginary part.

(c) Let $s=x>0$  be real and positive.  Then the first branch 
$f_1(s)=\sqrt{r_1\,r_2}\,e^{i\, 0}=\sqrt{r_1\,r_2}$ is also real and positive, while 
the second branch
$f_2(s)=\sqrt{r_1\,r_2}\,e^{i\, \pi}=-\sqrt{r_1\,r_2}$
is real and negative.

(d) Let $s=-x<0$ be real and negative.  Then 
the first branch  
$f_1(s)=\sqrt{r_1\,r_2}\,e^{-i\,\pi}=-\sqrt{r_1\,r_2}$ is real and negative, while the second branch 
$f_2(s)=\sqrt{r_1\,r_2}\,e^{i\, 0}=\sqrt{r_1\,r_2}$
is real and positive.

With these preparations done, let us return to 
the  function $\tilde R(s)$ given by Eq. (\ref{B_r1}). It 
has two branch points $\pm i\omega_0$ and four simple poles.
Two of the poles are on the imaginary axes
\begin{eqnarray}
 s=\pm i\,\omega_*, \qquad
 \omega_*=\sqrt{
 \frac{1+\sqrt{1+\alpha^2}}{2}
 }\,\omega_0>\omega_0,
\label{poles_imaginary}
\end{eqnarray}
and another two are on the real axis  
\begin{eqnarray}
 s=\pm c\,\omega_0, \qquad
 c=\sqrt{\frac{\sqrt{1+\alpha^2}-1}{2}}.
 \label{poles_real}
\end{eqnarray}
Let us show that the pure imaginary poles (\ref{poles_imaginary}) 
are on the first branch $\tilde R_1(s)$, and real poles (\ref{poles_real}) are on the second branch $\tilde R_2(s)$.
According to (\ref{B_r1}), 
each pole is a root
of the equation
\begin{eqnarray}
 \sqrt{s^2+\omega_0^2}=f(s)=-\frac{\alpha\,\omega_0^2}{2\,s}
\label{pole_eq}
\end{eqnarray}
for one of the two branches of the function $f(s)$. 
Let us determine for  each pole the corresponding branch of $f$ and $R$.

(1) At the pole $s=i\,\omega_*$ Eq. (\ref{pole_eq}) gives 
for $f(s)$ a pure imaginary value with 
a positive imaginary part. 
According to mapping rule (a), in this case $f(s)$ must be represented by the branch $f_1(s)$. Therefore, 
the pole is on the resolvent's first branch $\tilde R_1(s)$
for any  $\alpha$.

(2) At the pole $s=-i\,\omega_*$ Eq. (\ref{pole_eq}) gives 
for $f(s)$ a pure imaginary  value
with a negative imaginary part.
According to mapping rule (b), in this case $f(s)$ must be again represented by the branch $f_1(s)$, and  
the pole is on the resolvent's first branch $\tilde R_1(s)$
for any  $\alpha$.

(3) At the  pole $s=c\,\omega_0$ 
Eq. (\ref{pole_eq}) gives 
for $f(s)$ a real negative value.
According to mapping rule (c),
in this case $f(s)=f_2(s)$. The pole
is on the second branch $\tilde R_2(s)$
for any $\alpha$.

(4) At the pole $s=-c\,\omega_0$ 
Eq. (\ref{pole_eq}) gives 
for $f(s)$ a real positive value.
According to mapping rule (d), $f(s)=f_2(s)$.
The pole is 
on  the second branch $\tilde R_2(s)$.

As the  next step, we need to determine which of the two branches of
the function $\tilde R(s)$ is physically meaningful.
Interestingly, this task is more involved compared 
to the case $\beta=1$.  Consider, for instance,  
the condition $R(0)=1$. Using the initial value theorem it can be 
written as
\begin{eqnarray}
R(0)=\lim_{s\to\infty} s\,\tilde R(s)=\lim_{s\to\infty}
\frac{s\,f_k(s)}{s\,f_k(s)+\alpha\,\omega_0^2/2}=1.
\end{eqnarray}
One observes that this asymptotic relation is valid for both branches of
$f(s)$ and $\tilde R(s)$ as $s$ goes to infinity along any directions of both
real and imaginary axes of the complex plane.
Instead of the initial condition for the resolvent, we can use that for the memory kernel, $K(0)=\alpha\,\omega_0^2/2$,
see Eq. (\ref{K_initial}). For $\beta=2$ the transform of the kernel is given by Eq. (\ref{K2}),
\begin{eqnarray}
\tilde K(s)=\frac{\alpha\,\omega_0^2/2}{\sqrt{s^2+\omega_0^2}}.
\end{eqnarray}
Then the initial value theorem requires
\begin{eqnarray}
  K(0)=\lim_{s\to\infty} s\tilde K(s)=
  \frac{\alpha\,\omega_0^2}{2} \lim_{s\to\infty} \frac{s}{\sqrt{s^2+\omega_0^2}}=\frac{\alpha\,\omega_0^2}{2}.
\label{C_cond}
\end{eqnarray}
Suppose $s$ goes to infinity, say,  along the positive direction of the real axis.
Then, according to the mapping rules (c), condition (\ref{C_cond}) 
is only satisfied 
if the function $f(s)=\sqrt{s^2+\omega_0^2}$
is represented by its first branch $f_1(s)$.
The same conclusion we arrive at when  $s$ goes to zero along other directions.
Thus, the physical branch of $\tilde K(s)$
is the one involving the first branch $\tilde f_1$ of the square-root function $\tilde f(s)$. Since $\tilde R=1/(s+\tilde K)$,
the same  is true about the resolvent.
Therefore, the resolvent has to be found  
as the inversion of $\tilde R_1(s)$, i.e. as a Bromwich integral
\begin{eqnarray}
 R(t)=\frac{1}{2\pi i} \int_{\gamma-i\infty}^{\gamma+i\infty} e^{st}\, \tilde R_1(s)\,ds.
 \label{BromwichA}
\end{eqnarray}

As discussed above, $\tilde R_1(s)$ has two branch points $\pm i\,\omega_0$ and two simple poles $\pm i\,\omega_0$  
with $\omega_*>\omega_0$.
Since all four singular points are on the imaginary axis, the integral (\ref{BromwichA})
is over an arbitrary vertical line $s=\gamma$ 
to the right of the origin ($\gamma$ is real and positive).

\begin{figure*}[t]
\includegraphics[height=7cm]{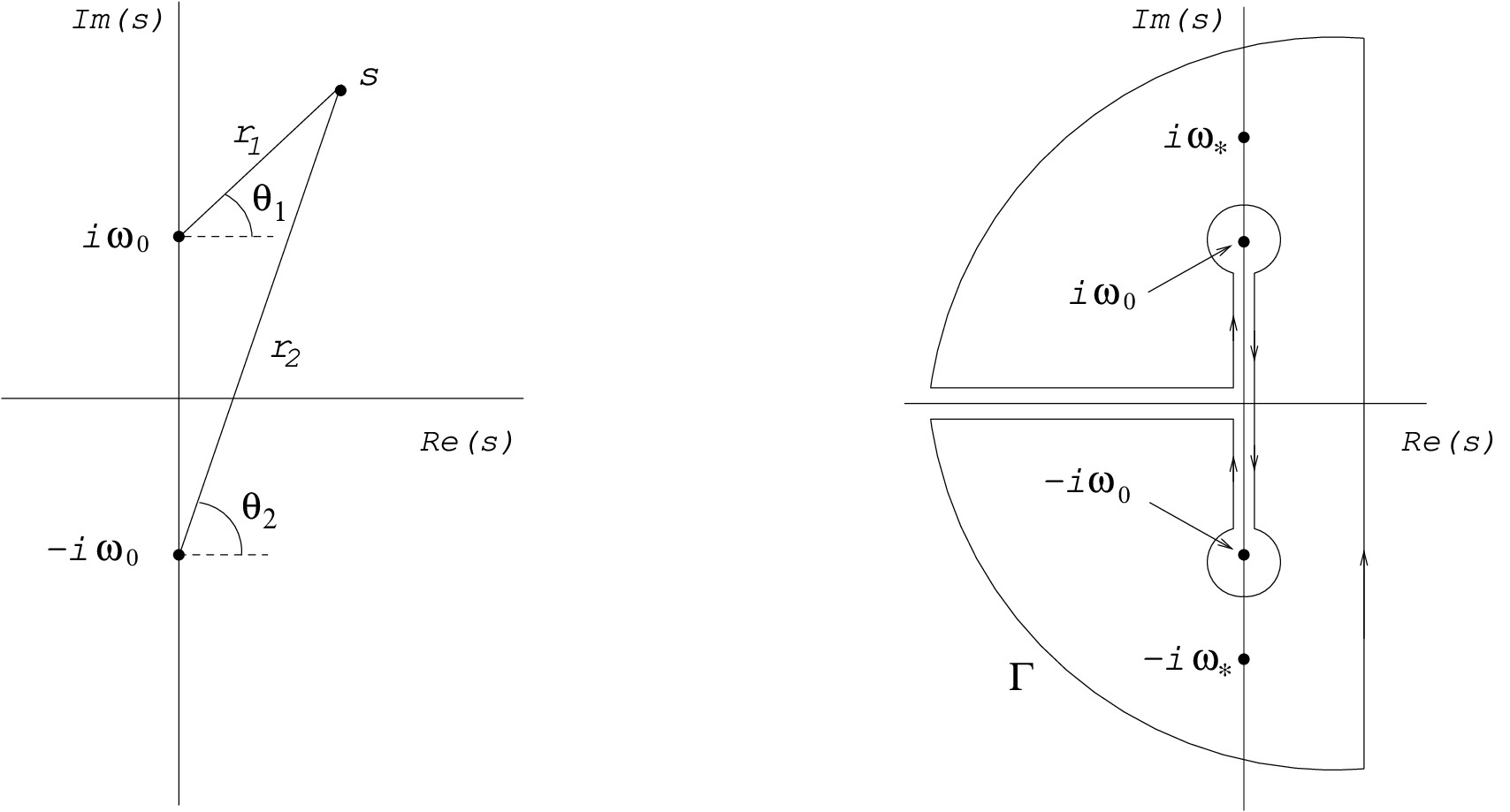}
\caption{Left: The definition of polar coordinates $(r_1,\theta_1)$ and $(r_2,\theta_2)$ in Eq. (\ref{polar}).
  Right: The integration contour  $\Gamma$ in the integral (\ref{integral_aux}).}
\end{figure*}

The evaluation of  integral (\ref{BromwichA})
is an exercise of the standard technique based on Cauchy's residue theorem. First, consider the auxiliary integral 
\begin{eqnarray}
 I(t)=\frac{1}{2\pi i} \int_\Gamma e^{st}\, \tilde R_1(s)\,ds=\mbox{Res}\,[e^{st}\tilde R_1(s),i\,\omega_*]
 +\mbox{Res}\,[e^{st}\tilde R_1(s), -i\,\omega_*]
 \label{integral_aux}
\end{eqnarray}
over the closed contour $\Gamma$ shown at the right part of Fig. 4. Here we use the notation $\mbox{Res}[f(z),z_0]$
for a residue of a function $f(z)$ at $z=z_0$.
One can show that contributions to the integral $I$ from the large arc (of radius $r$) and small
circles (of radius $\epsilon$) about branch points both go to zero when $r\to\infty$ and $\epsilon\to 0$.
The contribution from the two horizontal lines
along the negative real axis is also zero
when the distance between the lines vanishes because 
the integrand is continuous on the $x$-axis.
The only non-zero contributions to $I$ are those from the two vertical segments along the branch cut ($I_1$)
and from the vertical segment of the length $2r$ on the right ($I_2$), $I=I_1+I_2$. In the limit $r\to \infty $
the integral $I_2$ equals $R(t)$, therefore $I=I_1+R(t)$, and
\begin{eqnarray}
 R(t)=I(t)-I_1(t)=\mbox{Res}\,[e^{st}\tilde R_1(s), i\,\omega_*]
 +\mbox{Res}\,[e^{st}\tilde R_1(s), -i\,\omega_*]-I_1(t).
 \label{CA2}
\end{eqnarray}

The integral $I_1$ has two contributions
$I_1=I_1^-+I_1^+$. Consider first the contribution $I_1^-$ from the vertical path just left from the branch  cut, 
i.e. from $-i\,\omega_0-\epsilon$ to $i\,\omega_0-\epsilon$.
Using the path parametrization 
\begin{eqnarray}
 s(y)=i\,y-\epsilon, \qquad  -\omega_0<y<\omega_0
 \label{path1}
\end{eqnarray}
we can write $I_1^-$ in the form 
\begin{eqnarray}
 I_1^-=\frac{1}{2\pi i} \int_{-\omega_0}^{\omega_0}
e^{st}\,\tilde R_1(s)\,s'(y)\,dy=
\frac{1}{2\pi} \int_{-\omega_0}^{\omega_0}
e^{st}\,\frac{f_1(s)}{s\,f_1(s)+\alpha\,\omega_0^2/2}\,dy.
\label{I-}
\end{eqnarray}
As follows from Eq. (\ref{branch1}),  
on the given path
for the first branch $\theta_1=-\pi/2-\epsilon$ and 
$\theta_2=-3\pi/2+\epsilon$, and therefore 
\begin{eqnarray}
f_1(s)=\sqrt{r_1\,r_2}\,e^{i\frac{\theta_2+\theta_2}{2}}=
\sqrt{r_1\,r_2}\,e^{-i\, \pi}=-\sqrt{r_1\,r_2}.
\end{eqnarray}
Also,  it is easy to figure out that for the given path
$r_1\,r_2=(\omega_0-y)(\omega_0+y)$.
Then
\begin{eqnarray}
  f_1(s)=-\sqrt{r_1\,r_2}=-\sqrt{(\omega_0-y)\,(\omega_0+y)}=
  -\sqrt{\omega_0^2-y^2}.
\label{f1_path1}
\end{eqnarray}
Then integral $I_1^-$ takes the form
\begin{eqnarray}
 I_1^-=
-\frac{1}{2\pi} \int_{-\omega_0}^{\omega_0}
e^{iyt}\,\frac{\sqrt{\omega_0^2-y^2}}{\alpha\,\omega_0^2/2
-i\,y\,\sqrt{\omega_0^2-y^2}}\,dy.
\end{eqnarray}
Separating real and imaginary parts of the fraction yields 
\begin{eqnarray}
 I_1^-=
-\frac{\alpha\,\omega_0^2}{4\pi} \int_{-\omega_0}^{\omega_0}
e^{iyt}\,\frac{\sqrt{\omega_0^2-y^2}}{\alpha^2\omega_0^4/4+y^2\,(\omega^2-y^2)}\,dy
-\frac{i}{2\pi} \int_{-\omega_0}^{\omega_0}
e^{iyt}\,\frac{y\,(\omega_0^2-y^2)}{\alpha^2\omega^4/4+y^2\,(\omega^2-y^2)}\,dy.
\label{I-2}
\end{eqnarray}

In a similar way, using the path parametrization 
\begin{eqnarray}
 s(y)=i\,y+\epsilon, \qquad  -\omega_0<y<\omega_0,
 \label{path2}
\end{eqnarray}
one evaluates 
the second contribution $I_1^+$ from the vertical path just right from the branch  cut, 
\begin{eqnarray}
 I_1^+=-
\frac{1}{2\pi} \int_{-\omega_0}^{\omega_0}
e^{st}\,\frac{f_1(s)}{\alpha\,\omega_0^2/2+s\,f_1(s)}\,dy.
\label{I+}
\end{eqnarray}
Here the negative sign reflects  that the path is directed downward.
According to Eq. (\ref{branch1}),  
on the given path
for the first branch $\theta_1=-\pi/2+\epsilon$ and 
$\theta_2=\pi/2-\epsilon$, therefore 
\begin{eqnarray}
f_1(s)=\sqrt{r_1\,r_2}\,e^{i\frac{\theta_2+\theta_2}{2}}=\sqrt{r_1\,r_2}=
\sqrt{(\omega_0-y)\,(\omega_0+y)}=\sqrt{\omega_0^2-y^2},
\label{f1_path2}
\end{eqnarray}
and 
\begin{eqnarray}
 I_1^+=
-\frac{1}{2\pi} \int_{-\omega_0}^{\omega_0}
e^{iyt}\,\frac{\sqrt{\omega_0^2-y^2}}{\alpha\,\omega_0^2/2+iy\,\sqrt{\omega_0^2-y^2}}\,dy.
\end{eqnarray}
As for $I_1^-$, it is convenient to separate real and imaginary parts of the fraction, 
\begin{eqnarray}
 I_1^+=
-\frac{\alpha\,\omega_0^2}{4\pi} \int_{-\omega_0}^{\omega_0}
e^{iyt}\,\frac{\sqrt{\omega_0^2-y^2}}{\alpha^2\omega_0^4/4+y^2\,(\omega_0^2-y^2)}\,dy
+\frac{i}{2\pi} \int_{-\omega_0}^{\omega_0}
e^{iyt}\,\frac{y\,(\omega_0^2-y^2)}{\alpha^2\omega^4/4+y^2\,(\omega^2-y^2)}\,dy.
\label{I+2}
\end{eqnarray}
Adding up Eqs. (\ref{I-2}) and (\ref{I+2}),
and taking into account that the contribution from the odd part of the integrand is zero, one finds
\begin{eqnarray}
 I_1(t)=
-\frac{\alpha\,\omega_0^2}{2\pi} \int_{-\omega_0}^{\omega_0}
\frac{\cos(y\,t)\,\sqrt{\omega_0^2-y^2}}{\alpha^2\omega_0^4/4+y^2\,(\omega_0^2-y^2)}\,dy=
-\frac{\alpha\,\omega_0^2}{\pi} \int_{0}^{\omega_0}
\frac{\cos(y\,t)\,\sqrt{\omega_0^2-y^2}}{\alpha^2\omega_0^4/4+y^2\,(\omega_0^2-y^2)}\,dy.
\label{I12}
\end{eqnarray}

The next step is to evaluate the residues in  expression (\ref{CA2}).
One can verify that the poles are of the first order, then
\begin{eqnarray}
\mbox{Res}\,[e^{st} \tilde R_1(s), i\,\omega_*]=\lim_{s\to i\,\omega_*}
e^{st}\,\tilde R_1(s)\,(s-i\,\omega_*)=e^{i\,\omega_* t}
\lim_{s\to i\,\omega_*} \frac{f_1(s)\,(s-i\,\omega_*)}{s\,f_1(s)+\alpha\,\omega_0^2/2}.
\end{eqnarray}
Using the L'Hospital's rule one gets
\begin{eqnarray}
\mbox{Res}\,[e^{st} \tilde R_1(s), i\,\omega_*]=e^{i\,\omega_* t}
\lim_{s\to i\,\omega_*} \frac{f_1^2(s)}{s^2+f_1^2(s)}.
\label{residues}
\end{eqnarray}
According to mapping rule (a)
\begin{eqnarray}
f_1(i\,\omega_*)=i\,\sqrt{r_1\,r_2}=i\,
\sqrt{(\omega_*-\omega_0)(\omega_*+\omega_0)}=i\,\sqrt{\omega_*^2-\omega_0^2}.
\end{eqnarray}
Then 
\begin{eqnarray}
\mbox{Res}\,[e^{st} \tilde R_1(s), i\,\omega_*]=e^{i\,\omega_* t}\,
\frac{(\omega_*/\omega_0)^2-1}{2\,(\omega_*/\omega_0)^2-1}.
\label{residues1}
\end{eqnarray}
The second pole at $-i\,\omega_*$ is evaluated in a similar way, so we get
\begin{eqnarray}
\mbox{Res}\,[e^{st} \tilde R_1(s), \pm i\,\omega_*]=e^{\pm i\,\omega_* t}\,
\frac{(\omega_*/\omega_0)^2-1}{2\,(\omega_*/\omega_0)^2-1}=\frac{e^{\pm i\,\omega_* t}}{2}\,
\left(
1-\frac{1}{\sqrt{1+\alpha^2}}
\right).
\label{residues2}
\end{eqnarray}

Finally, the substitution of expressions (\ref{I12}) for $I_1$ and
(\ref{residues2}) for the residues into Eq. (\ref{CA2}) yields
\begin{eqnarray}
R(t)=R_0(t)+A\, \cos \omega_* t.
\label{R_final}
\end{eqnarray}
Here the term
\begin{eqnarray}
R_0(t)=-I_1(t)=
\frac{\alpha\,\omega_0^2}{\pi} \int_{0}^{\omega_0}
\frac{\cos(y\,t)\,\sqrt{\omega_0^2-y^2}}{\alpha^2\omega_0^4/4+y^2\,(\omega_0^2-y^2)}\,dy.
\end{eqnarray}
can be shown to vanish in the limit $t\to \infty$, and the amplitude and frequency of the 
oscillatory term are 
\begin{eqnarray}
A=1-\frac{1}{\sqrt{1+\alpha^2}}, \qquad 
\omega_*=\sqrt{
 \frac{1+\sqrt{1+\alpha^2}}{2}
 }\,\omega_0.
\end{eqnarray}
Although it is not immediately obvious, one can verify numerically that
the result
(\ref{R_final}) satisfies the correct initial condition $R(0)=1$.

\end{document}